\documentclass[aps,prx,twocolumn,showpacs,superscriptaddress,groupedaddress]{revtex4}

\usepackage[dvips]{graphicx}
\usepackage{amssymb,amsfonts,amsmath}
\usepackage{eucal,bm}
\usepackage{gensymb,subfigure}
\usepackage{color,url}

\begin{document}

\title{Cascading of Fluctuations in Interdependent Energy Infrastructures:\\
Gas-Grid Coupling}

\author{Michael Chertkov}
\affiliation{Theoretical Division \& Center for Nonlinear Studies, Los Alamos National Laboratory, Los Alamos, NM 87545, USA}

\author{Vladimir Lebedev}
\affiliation{Landau Institute for Theoretical Physics, Moscow, Russia}

\author{Scott Backhaus}
\affiliation{Materials Physics and Applications Division,\\ Los Alamos National Laboratory, Los Alamos, NM 87545, USA}

\begin{abstract}
The revolution of hydraulic fracturing \cite{fracking} has dramatically increased the supply and lowered the cost of natural gas in the United States driving an expansion of natural gas-fired generation capacity in many electrical grids \cite{ISO-NE}.  Unrelated to the natural gas expansion, lower capital costs \cite{cost_renewables} and renewable portfolio \cite{portfolio_renewables} standards are driving an expansion of intermittent renewable generation capacity such as wind and photovoltaic generation.  These two changes may potentially combine to create new threats to the reliability of these interdependent energy infrastructures.  Natural gas-fired generators are often used to balance the fluctuating output of wind generation.  However, the time-varying output of these generators results in time-varying natural gas burn rates that impact the pressure in interstate transmission pipelines. Fluctuating pressure impacts the reliability of natural gas deliveries to those same generators and the safety of pipeline operations.  We adopt a partial differential equation model of natural gas pipelines and use this model to explore the effect of intermittent wind generation on the fluctuations of pressure in natural gas pipelines.  The mean square pressure fluctuations are found to grow linearly in time with points of maximum deviation occurring at the locations of flow reversals.
\end{abstract}

\maketitle

%\keywords{Natural Gas Network | Power Grid Network | Optimization | Uncertainty | Fluctuations}

%\abbreviations{ISO-NE, Independent System Operator of New England; PV, Photo-Voltaic; LDC, Load Distribution Companies; PDE, Partial Differential Equations }

%\section{Significance Statement}

The ongoing evolution to intermittent wind and solar electric generation is causing many electrical grid operators to use more agile natural gas-fired electric generation to balance these new stochastic resources. This interdependence causes a cascade of the fluctuations of renewable generation into the systems that supply fuel to the gas-fired generators, i.e. natural gas pipelines. We develop a model of the coupling between electrical grid fluctuations and natural gas pipeline systems, analyze the resulting fluctuations of pipeline pressure, and draw conclusions about the impact of renewable electrical generation on the stability and security of natural gas pipelines.

\section{Introduction}

\begin{figure}
\begin{center}
\includegraphics[width = 2.5 in]{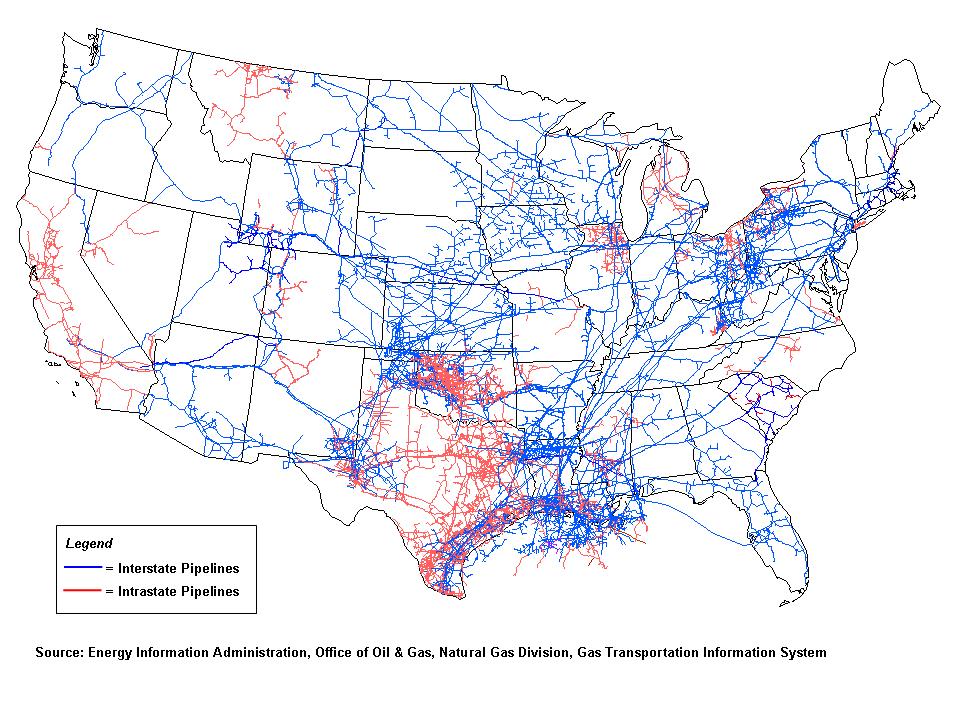}
\includegraphics[width = 2.5 in]{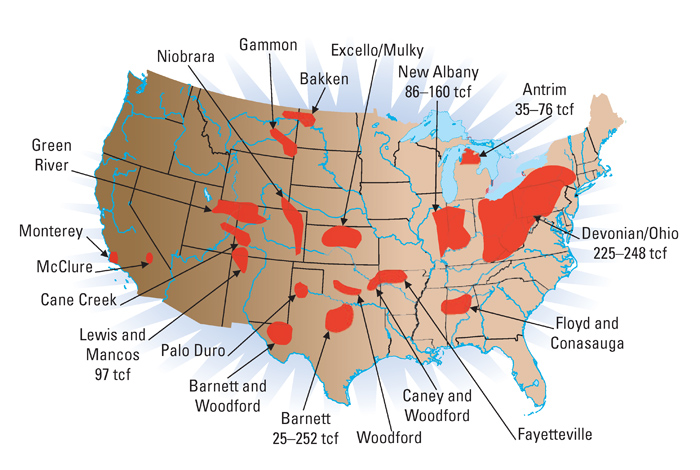}
\end{center}
\caption{ (Left) The Natural Gas Pipeline Network of the United States. Interstate pipelines are not significantly meshed and primarily display a tree-like structure. (Right) Major US shale gas basins -- new sources of natural gas that will encourage realignment of US National Gas Network.}
\label{fig:gas_pipes}
\end{figure}

By making unconventional natural gas sources economic to extract, hydrofracking has created a revolution in the U.S. natural gas industry \cite{fracking}.  Many of these new gas sources are in nontraditional locations such as the Marcellus shale in Pennsylvania, the Niobrara shale in Eastern Colorado, and the Bakken shale in North Dakota. See Fig.~\ref{fig:gas_pipes}. The dramatic increase in supply has driven down prices and spurred many new or expanded uses for natural gas \cite{2010MITEI,2013MITEI}.  This revolution in the natural gas supply and loads is creating new challenges for natural gas pipelines that transport the gas from source to load.  With a limited amount of throughput and short-term gas storage (in the form of pressure in pipeline itself), these pipelines may become vulnerable as their operating environment changes.

\begin{figure}
\begin{center}
\includegraphics[width = 2.5 in]{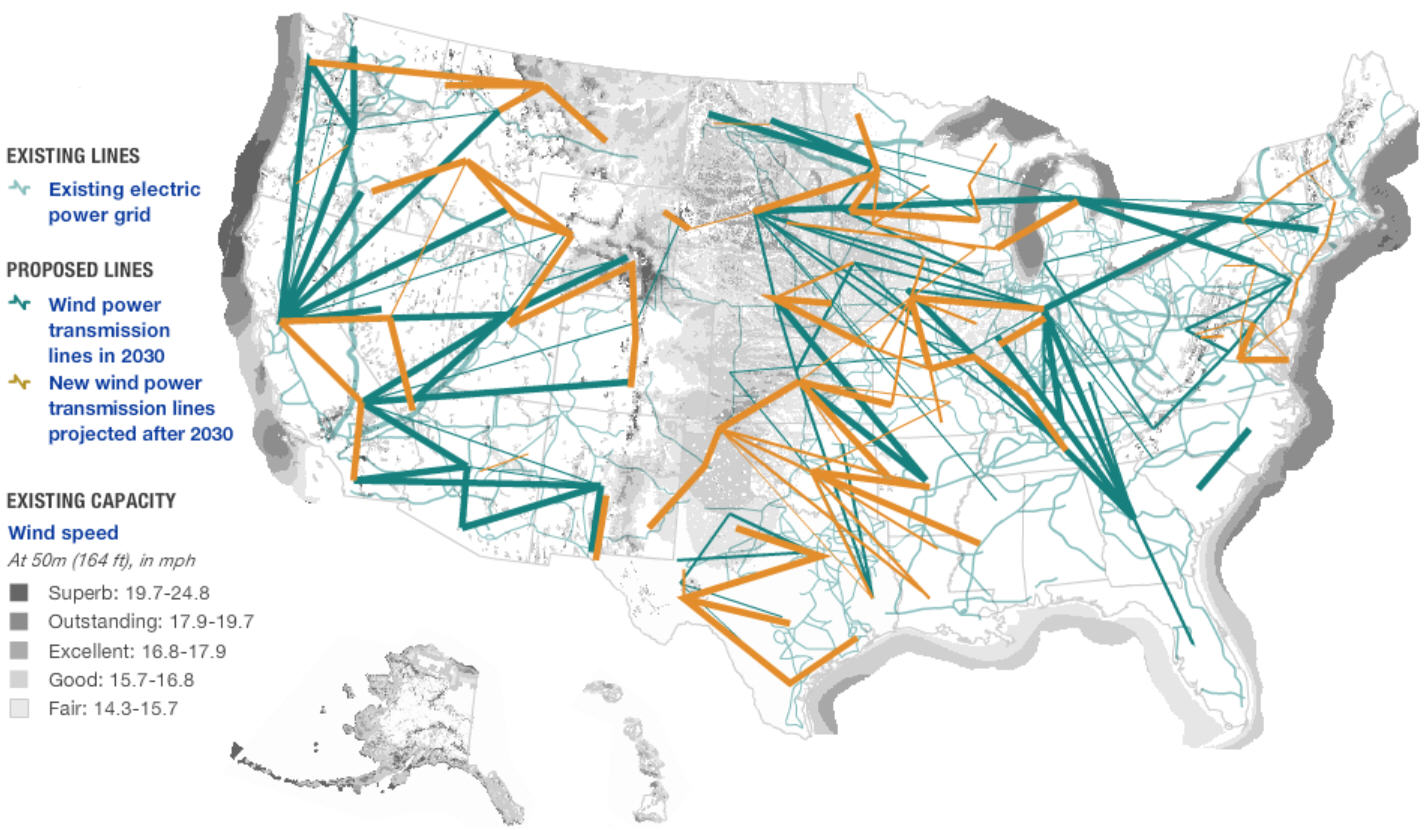}
\includegraphics[width = 2.5 in]{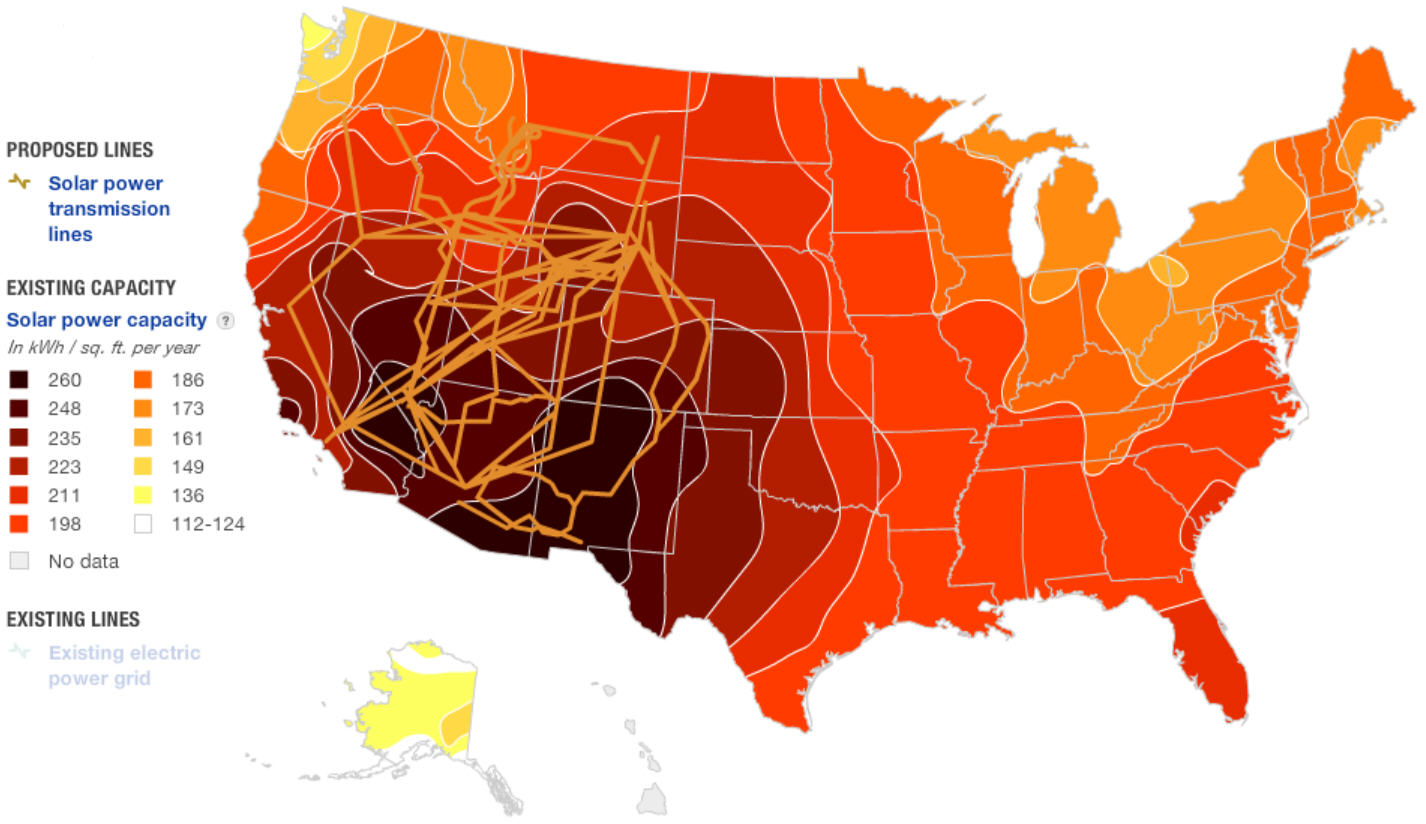}
\end{center}
\caption{ (Left) US Power Transmission Grid (including potential future transmission expansions) superimposed on wind power capacity map. (Right) Solar power capacity map with proposed transmission lines to improve the integration of solar resources into the existing power grid. (Adapted from National Public Radio, Visualizing the U.S. Electric Grid, 2009.)
\label{fig:power_transmission}
}
\end{figure}

A dominant new load on the gas pipelines is natural gas-fired generators.  Previously, the marginal cost of electricity from these generators was higher than from coal-fired generators. However, the rapid drop in gas prices has made gas generation competitive with coal and spurred its construction.  An example of this dramatic expansion is in the electrical grid controlled by the Independent System Operator of New England (ISO-NE) where natural gas-fired electrical generation increased from 5\% of total capacity to 51\% in a span of 20 years \cite{ISO-NE}.  A parallel development in many U.S. electrical grids is the expansion of intermittent renewable generation such as wind and PhotoVoltaic (PV) generation---a trend that is expected to continue as utilities work to meet state-imposed renewable portfolio standards that mandate a certain fraction of electrical generation be derived from renewable sources. See Fig.~\ref{fig:power_transmission}. In contrast to traditional nuclear, coal, or gas-fired generation, these new forms of generation have a small degree of controllability.  To maintain the second-by-second balance of generation and load, other grid resources must respond to counteract the fluctuations of the intermittent generation.  Although many different types of advanced control of nontraditional resources are being considered to provide these balancing services, e.g. grid-scale battery storage and demand response, the currently most available resources are the controllable traditional generators with gas-fired generators being the most flexible among these.

The combination of expanded natural gas-fired generation and its increased use to balance intermittent renewable generation is creating loads on natural gas pipelines that are significantly different than in the past.  Traditional gas pipeline loads (Load Distribution Companies or LDCs) primarily serve space or water heating or other individual customer needs and evolve slowly throughout the day in a relatively well-known pattern that can be predicted based on historical information and weather forecasts. Other traditional pipeline customers are industrial loads that, although they may change from day to day, are very predictable over the span of a day.  In contrast, when gas-fired generation is used to balance fluctuating renewable generation, a component of the resulting gas loads take on a stochastic nature. Unlike the gas load of an LDC, wind and PV generation respond to {\it short-term} fluctuations in environmental conditions, e.g. wind fluctuations on the timescale of 10-100 minutes and solar insolation fluctuations on the timescale of 1-100 minutes. At the longer timescales, these fluctuations may contain spatiotemporal correlations that increase the aggregate fluctuations of wind or PV generation across an entire electrical grid magnifying the fluctuations of natural gas loads used by gas generators to balance these changes.

Fluctuating gas loads create new dynamics in natural gas pipelines that can impact their reliability and the reliability of all interdependent infrastructures, including the electrical grid.  To a great extent, electrical grid dynamics are determined by the very small amount of energy stored in the rotating kinetic energy of large centralized generators.  Under a serious upset, this energy storage can maintain the reliable operation of the grid for a second or two while other resources are adjusted to compensate---typically an adjustment of mechanical power input to these same generators supported by a change in fuel burn rate.  In some respects, if the grid ``storage'' is sufficient to survive the initial upset, an electrical grid with fuel-burning generators has very large amounts of storage on longer timescale in the fuel supplied to those generators.

Gas pipelines dynamics evolve on very different timescales.  In the short term (10-100 minutes), gas pipelines have a large amount of storage in the compressed natural gas in the pipeline itself. A typical gas pipeline might be run very near its upper limit on pressure of 800 psi whereas the minimum gas delivery pressure is 500 psi.   Even if all gas injections into the pipeline were cutoff, the gas loads would slowly reduce the pressure of compressed gas over a few hours without any significant impact on the loads until the gas pressure falls below the minimum delivery pressure.  However, unlike the electrical grid, there may be very little gas storage on longer timescales.  Injections of gas into the pipeline are scheduled via bi-lateral transactions in gas markets and are typically held constant throughout a 24-hour period.  Therefore, the gas pressure in the pipeline evolves over both space and time according to the spatiotemporal arrangements of the gas injection and gas loads.  If the injections and loads are out of balance, the gas pressure will undergo spatiotemporal evolution.  However, the fluctuations in pressure will not be spatially uniform.  In fact, the pressure fluctuations are nonlocal with the swings at one locations depending on the behavior at all other locations.

The feedback between fluctuating gas loads and gas pressure creates coupled reliability concerns across the natural gas pipeline and electrical grid infrastructures.  The nonlocal effects mentioned above can couple with spatiotemporal correlations in the fluctuations of renewable generation through the response of the gas generators to magnify pressure fluctuations at certain locations in the pipeline.  These fluctuations may lead to significant over or under pressures, both of which have serious impact on the reliability and safety of the pipeline itself.  Under pressures may impact the gas generators by forcing them to reduce electrical output or potentially shutdown to preserve the integrity of service to other pipeline customers.  As we will show in this manuscript, the most sensitive locations are those of zero flow at the end of the pipeline with unidirectional flow or at location(s) of flow reversal in pipelines with well separated injection locations.  Therefore, the evolving spatial dependence of U.S. natural gas supply will couple to the stochasticity to create additional uncertainty in the reliability of the gas and electrical systems.

Neither gas pipeline nor electrical grid operators have the analysis tools to sufficiently address the probabilistic nature of the reliability impacts created by the coupled stochasticity of these infrastructures. The goal of the manuscript is to lay the foundation for these tools by developing a model and analysis to predict the spatiotemporal evolution of the probability distribution of gas pipeline pressure fluctuations.  This first step seeks to develop a measure of probabilistic risk that can be subsequently integrated into the operations of both the electrical and pipeline infrastructure networks.  We approach the problem by adopting phenomenological gas flow equations consisting of Partial Differential Equations (PDE) in one spatial dimension that have been accepted as accurate representations of long natural gas pipelines \cite{osiadacz1987simulation,87TT,05Sar}.

We develop models of fluctuations of gas-fired generator natural gas loads and the constraints imposed by natural gas markets to analyze the stochastically-driven PDEs.  We develop analytic expression for probability distributions of gas pipeline pressures as a function of space and time and as a function of the settings of gas compressor stations that push the gas along the pipeline. Our analysis shows that, under constant compressor station settings, the mean square pressure fluctuations grow linearly in time similar to a diffusive process.  We find that the largest mean square pressure fluctuations occur at location of zero flow that can potentially occur at any location along the pipeline depending on the average natural gas injections and loads. The results form the basis for a risk-aware optimization problem for the gas compressor stations controls.

The material in the rest of the manuscript is organized as follows.  Section \ref{sec:model} describes the basic model of natural gas pipelines. Section \ref{sec:fluct} describes pressure sensitivity to fluctuating gas draws. Future work and extensions are discussed in Section \ref{sec:conclusions}. Appendixes describe in greater detail the physical models
of gas flow and the approximations used to develop the models discussed in the main text.

\section{Model of Natural Gas Pipelines}\label{sec:model}

\begin{figure}
\centering
\includegraphics[width=0.45\textwidth]{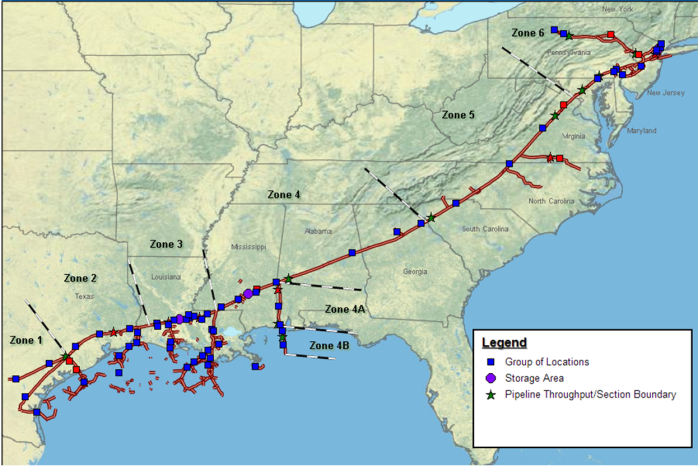}
\caption{Schematic representation of the Transco gas transmission network.}
\label{fig:Transco}
\end{figure}

The Transco pipeline (see Fig.~\ref{fig:Transco}) is a major interstate pipeline that delivers large quantities of natural gas to population centers and to natural gas-fired generators that supply electricity to those same population centers.  Like many other major interstate pipelines, the Transco pipeline displays a nearly radial structure and it is equipped with many compressors that are often nearly equally spaced along its length ($\sim$50-100 km between compressors).  These two properties are reasonably well approximated by the radial, distributed compression model discussed in the main text. Under these two approximations, the Transco and similar pipelines can be analyzed using the simplified models and analysis discussed in the following.

We adopt a phenomenological, spatially one-dimensional model of a transmission pipeline
delivering gas over long distances ($\sim$ 1000's of km)---a reasonable model of interstate pipelines in the US. The form of this model is generally accepted as an accurate representation of long pipelines \cite{osiadacz1987simulation,87TT,05Sar}. See the supplementary information (Appendixes) for model derivation and additional details. The gas injections may be configured in many different ways, e.g. at a single source at the originating end of the pipeline, two sources at either end of the pipeline, or in a distributed manner along the pipeline.  However, in all that follows, the injections will be assumed to be constant in time--a simplification that is also a close representation of pipeline operations in the U.S.  Natural gas loads are distributed along the pipeline and may fluctuate in time. Pressure gradients drive the gas along the pipeline from sources to sinks, and these gradients are maintained by gas compressors. A few other assumptions in the derivation and analysis of the model are made, but these are mostly taken to simplify the presentation.  We will point out where these assumptions can be removed via more complicated analysis.

\subsection{Gas Dynamic Equations}
%\underline{\bf Gas Dynamics Equations.}

By integrating over the cross section of the natural gas pipeline, the three-dimensional equations of hydrodynamics are reduced to a representation in one spatial dimension.   Mass conservation becomes
\begin{equation}\label{eq:mass_con}
 c_s^{-2}\partial_t p+\partial_x\phi=-q,
\end{equation}
where $t$ is time, $x$ is coordinate along the pipe ($0<x<L$), $p$ is the pressure along
the pipe, $\phi$ is the mass flow along the pipe, $q(x)$ is the density of the
distributed gas consumption ($q>0$ for injection and $q<0$ for loads), and $c_s$ is sound velocity of the gas. Using a friction factor $\beta$ as a  phenomenological representation of turbulent drag, Navier-Stokes equation becomes
\begin{equation}\label{eq:NS}
 \partial_x p+ \frac{\beta}{2d} \frac{\phi|\phi|}{p}=\gamma p.
\end{equation}
Here, $d$ is the pipe diameter and $\gamma(x)$ is a distributed representation of the many compressor stations in long pipelines. A real compressor station can operate in several different modes, one of which is a fixed compression ratio $\gamma$ such that $p_{out}=\gamma p_{in}$ where $p_{out}$ and $p_{in}$ are the pressures at the outlet and inlet of the compressor.  Here, we have distributed this lumped compression ratio along the pipeline such it makes a positive contribution to the $\partial_x p$ of size $\gamma(x)$.  Fast acoustic transients are ignored in Eq.~(\ref{eq:NS}) by assuming $c_s\gg u$, where $u$ is the typical gas velocity.  $u$ is generally small enough that this condition holds everywhere in the pipeline, however, $u$ (and its associated Reynolds number) is also large enough that $\beta$ can be taken to be constant. We note that Eqs.~(\ref{eq:mass_con},\ref{eq:NS}) have already been supplemented with an ideal gas equation state for the natural gas of the form $p=c_s^2 \rho$. The model derivation in the Appendixes addresses more general and more realistic settings such as meshed networks and compression spatially concentrated at the nodes.

\subsection{Simplified Market Model}
%\underline{\bf Simplified Market MOdel.}

The flow of natural gas in a pipeline is scheduled via bilateral transactions between buyers and sellers in a day-ahead market with market clearing and gas flows scheduling done well in advance of the following 24-hour period of gas delivery.  Scheduling consists of determining the locations and constant rates of gas injections.  The gas pipeline operator expects that gas loads will be fairly uniform over the 24-hour delivery period.  Some level of fluctuating gas load is allowed, and it is these fluctuations that is expected to grow as natural gas-fired electrical generation is increasingly used to balance renewable fluctuations.  After the 24-hour delivery period begins and gas buyers have better estimates of their actual needs, they can make mid course corrections by transacting and scheduling gas flows in two subsequent intra-day markets at 10 and 14 hours after the start of the 24-hour delivery period.  In the three intervening periods, the gas injections are held relatively constant, and it is these periods we seek to analyze.

We model these subperiods by first solving for a stationary solution where time-averaged gas loads $q^{(\mbox{st})}(x)$ are given and are globally balanced by time-independent gas injections at either end of the pipeline, i.e. $\phi^{(\mbox{st})}(0)=\phi_0\geq0$ and $\phi^{(\mbox{st})}(L)=\phi_L\leq0$.  The stationary gas flow along the pipe is related to the loads by
\begin{equation}\label{eq:stationary_eq}
\partial_x \phi^{(\mbox{st})}(x)=-q^{(\mbox{st})}(x),
\end{equation}
and the global mass balance implies
\begin{equation}\label{eq:stationary_balance}
\phi^{(\mbox{st})}(0)-\phi^{(\mbox{st})}(L)= \int_0^L dx\ q_0(x).
\end{equation}
Natural gas pipeline operators require that Eq.~(\ref{eq:stationary_balance}) be satisfied over the 24-hour delivery period.  To insure this condition over the 24-hour period, there may be some deviation in the balance in the intra-day periods to compensate for inaccurate forecasting or changes in the average gas loads.  However, in the remainder of this discussion, we will assume that the stationary solution is balanced in each intra-day subperiod.  In the following,
we will add fluctuations to $q^{(\mbox{st})}(x)$, and therefore $\phi^{(\mbox{st})}(x)$, to model the affects of renewable generation on the pipeline pressure fluctuations.

\subsection{Compressor Model For Stationary Flows}
%\underline{\bf Compressor Model for Stationary Flows.}

Before adding fluctuations, we first describe the control of the gas compressors for the stationary gas flows. If the gas loads $q(x)$ and flow $\phi(x)$ were actually stationary, then the control for the gas compression stations could be computed once and implemented for the entire 24-hour gas delivery period, or at least for the intra-day periods.  Natural gas pipeline operators seek to maintain a relatively uniform pressure profile up to the the pressure drop between compressor stations.  Our simple model of spatially-distributed compression $\gamma(x)$ in Eq.~(\ref{eq:NS}) is a reasonable representation of gas pipeline operations and provides a spatially uniform pressure $p_0$ when
\begin{equation}\label{eq:stationary_compression}
 \gamma(x)=\frac{\beta \phi^{(\mbox{st})}(x)|\phi^{(\mbox{st})}(x)|}{2d\, p_0^2}.
\end{equation}
We pick this model for ease of presentation.  Spatially discrete compression and nonuniform pressure profiles \cite{68WL,00WRBS,12BNV,13MFBBCP} can be incorporated in an edge-node network model in straightforward manner. See Appendixes for additional discussions.

\section{Pressure Sensitivity to Fluctuating Gas Draws}
\label{sec:fluct}

Time-dependent gas loads require the solution of the dynamic versions of Eqs.~(\ref{eq:mass_con},\ref{eq:NS}).  Here, we consider the time-dependent component to be fluctuations of the gas loads about their forecasted values, $q(t;x)=q^{(\mbox{st})}(x)+\xi(t;x)$, where $\xi(t;x)$ models the random, zero mean and statistically stationary fluctuations.  As described in Appendix \ref{sec:methods}
%the Method Section of the Supplemental Information (SI),
when these fluctuations are relatively weak (even though they may be spatio-temporarily nontrivial), %Eqs.~(\ref{eq:mass_con},\ref{eq:NS}) admit
an analytical solution for the time-dependent variance of the gas pressure %($\langle \delta p(t;x)^2 \rangle$).
valid at $t\gg\tau$ becomes
%For $t \gg \tau$, we may safely drop the $b$ in favor of $a(t)Z(x)$ and estimate the pressure variation covariance as
\begin{eqnarray}
&& \frac{\langle (\delta p(x))^2 \rangle}{p_0^2} =
\left(\frac{\overline{q}^{(\mbox{st})}c_s^2 \tau}{p_0^2}\right)^2 \frac{t}{\tau}\left(\frac{Z(x)}{Y}\right)^2
\nonumber \\ &&
\times \iint_0^L \frac{dx_1\ dx_2}{L^2} \frac{\langle \xi(t,x_1)\xi(t,x_2)\rangle}{(\overline{q}^{(\mbox{st})})^2}.
\label{eq:mean_square_nondim}
\end{eqnarray}
The solution shows that a pipeline's sensitivity to fluctuating gas draws depends on the stationary solutions $\phi^{(\mbox{st})}(x)$ or $\gamma(x)$ and the statistics of the fluctuating gas loads.

\begin{figure}
\begin{center}
\includegraphics[width = 3.5 in]{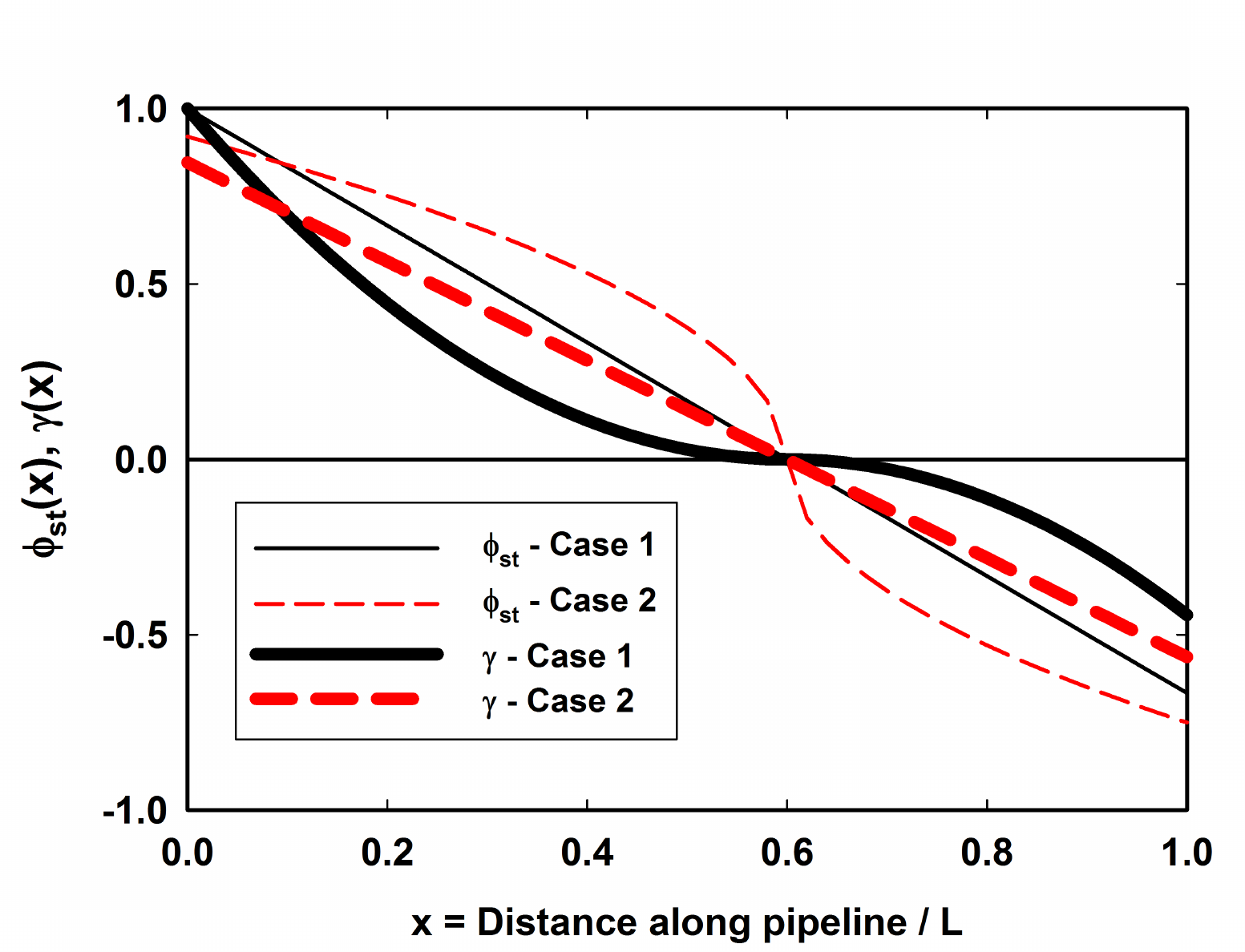}
\end{center}
\caption{Stationary mass flux $\phi^{(\mbox{st})}(x)$  and compression $\gamma(x)$ though the pipeline versus the distance $x$ along the pipeline.  Here, the length of the pipeline as been set to one, i.e. $L$= 1.  The plot shows two different cases of stationary mass flux to demonstrate the effect on the spatial dependence of the sensitivity parameter $Z(x)/Y$.  In both cases, the flow reversal occurs at $x^*$= 0.6.  The combined mass flux into the pipe from $x$= 0 and $x$= 1 is the same for both cases.  Case 1 is comprised of uniformly distributed gas loads at all locations along the pipeline.  Case 2 represents a combination of some distributed load along with a concentrated load at $x=x^*$=0.6.
}
\label{fig:Mass_flux}
\end{figure}

Here, we analyze three exemplary stationary configurations to explore the qualitative features of this sensitivity.  The first two cases are shown in Fig.~\ref{fig:Mass_flux}.
\begin{itemize}
\item Case 1 displays injection of gas only at the two ends of the pipeline ($\phi^{(\mbox{st})}(0)/\phi_0=1$ and $\phi^{(\mbox{st})}(L)/\phi_0=-2/3$) with uniformly distributed gas draws along the pipeline ($q^{(\mbox{st})}$=const).  The resulting mass flux along the pipeline is $\phi^{(\mbox{st})}(x) = \phi_0(1-x/x_*)$ -- thin black trace in Fig. \ref{fig:Mass_flux} -- where $x_*= 0.6L$.
    According to Eq.~(\ref{eq:stationary_compression}), the stationary compression is proportional to $\phi^{(\mbox{st})} |\phi^{(\mbox{st})}|$ creating a compression profile that goes to zero at $x_*$ and is biased toward either end of the pipeline at $x=0$ or $x=L$ -- thick black trace in Fig.~\ref{fig:Mass_flux}.

\item Case 2 is a simple modification of Case 1 that concentrates the gas draws near a single location at $x_*$.  In Case 2, $\phi^{(\mbox{st})}(x) = \phi_0\; 0.918\; \mathrm{sign}(1-x/x_*) \sqrt{|1-x/x_*|}$ -- thin red trace in Fig.~\ref{fig:Mass_flux} --
    where the choice of the $0.918$ factor for Case 2 makes {\it the total gas injection into the pipeline}, $\phi^{(\mbox{st})}(0)-\phi^{(\mbox{st})}(L)$, the same as in Case 1.  The larger gas draws near to $x_*$ are indicated by the higher values of $\partial_x \phi^{(\mbox{st})}$ at $x_*$.  The resulting compression shows a linear dependence with heavier weighting of compression closer to $x_*$ than in Case 1.
\end{itemize}

\begin{figure}
\begin{center}
\includegraphics[width = 3.5 in]{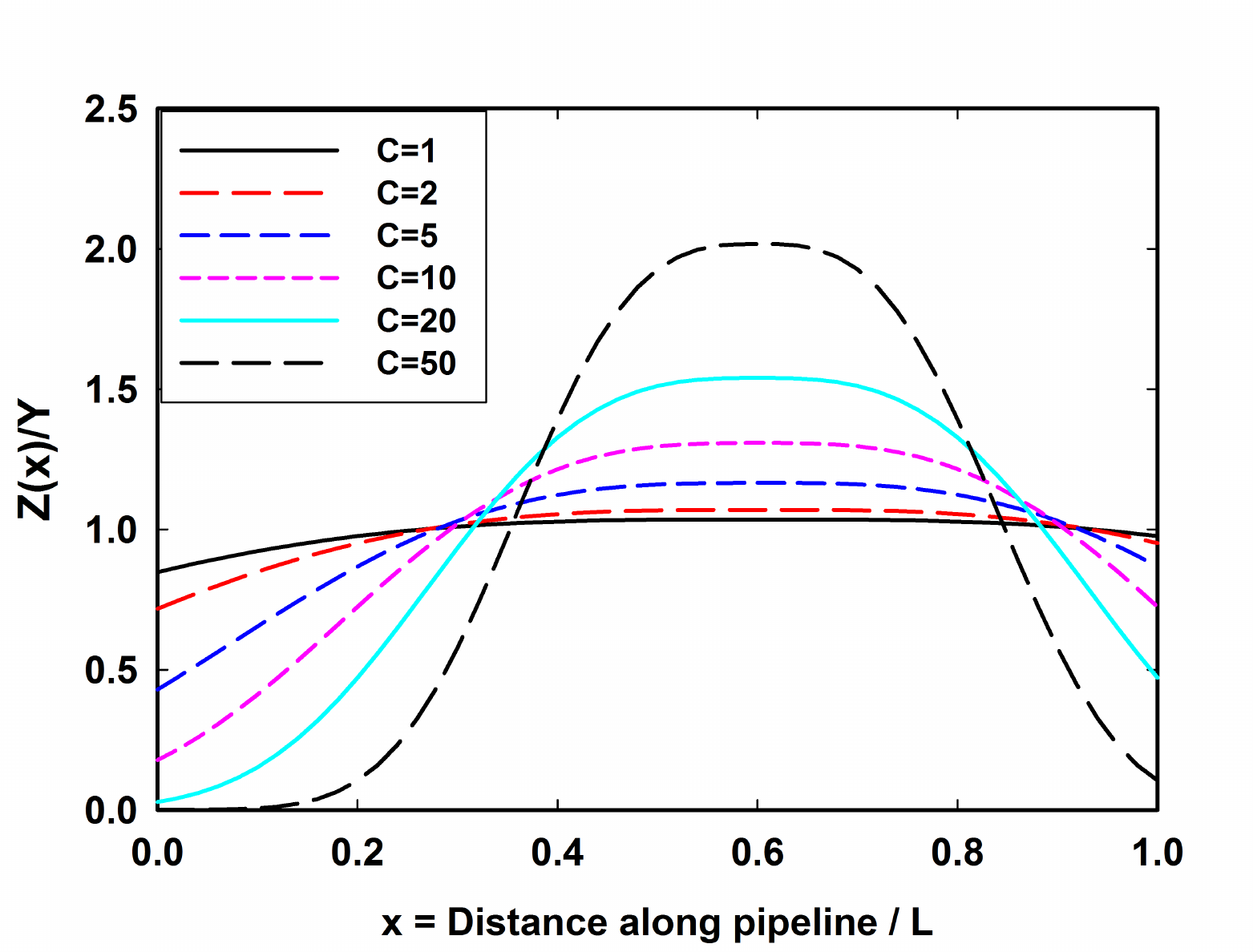}
%\vspace{-0.5cm}
\end{center}
\caption{The fluctuation sensitivity parameter $Z(x)/Y$ versus $x$ for Case 1 in Fig.~\ref{fig:Mass_flux}. For a given pipeline geometry, the different curves represent different scalings of the total stationary mass flux into the pipeline or the total compression deployed in the stationary solution.}
\label{fig:ZoverY_linear}
\end{figure}

\begin{figure}
\begin{center}
\includegraphics[width = 3.5 in]{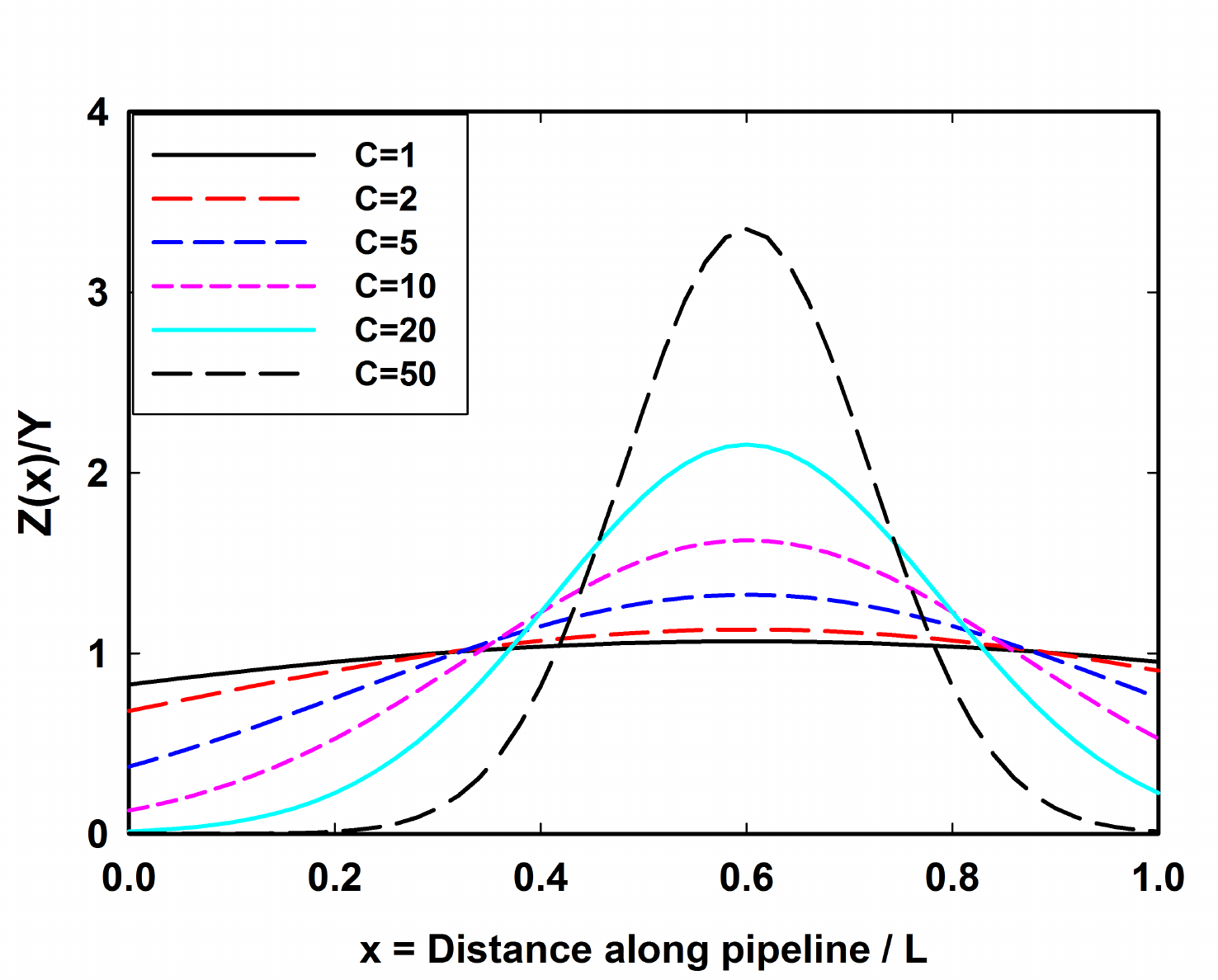}
\end{center}
\caption{Same as Fig.~\ref{fig:ZoverY_linear} except for Case 2 in Fig.~\ref{fig:Mass_flux}.}
\label{fig:ZoverY_sqrt}
\end{figure}

Using the distributed compression $\gamma(x)$ in Fig.~\ref{fig:Mass_flux},
$Z(x)/Y$ is computed using Eq.~(\ref{eq:mean_square_nondim}).  The results for Case 1 and Case 2 are shown in Fig.~\ref{fig:ZoverY_linear} and Fig.~\ref{fig:ZoverY_sqrt},
respectively.  The different traces in these Figures are for different values of the coefficient, $C\equiv \beta \phi_0^2 L/(d p_0^2)$, that scales the compression density $\gamma(x)$.  Interpreting the distributed compression in terms of a set of discrete compressor stations of uniform compression ratio, Case 2 with $C= 50$ corresponds to between 6 and 7 compressors with compression ratio 1.5 placed uniformly between $x$= 0 to $x_*$, i.e. a typical number of compressors for a mildly stressed pipeline configuration.

When the pipeline is under very little stress ($C= 1$), both Case 1 and Case 2 show a relatively uniform $Z(x)/Y\sim 1$.  Under these conditions, there are no regions of the pipeline that show a significantly enhanced sensitivity to stochastic gas loads.  As the stress is increased (larger $C$), proportionally more compression is deployed in the stationary solution.  Both Case 1 and Case 2 show a depression of $Z(x)/Y$ near the ends of the pipeline, i.e. the injection points, and an enhancement near $x_*$.  Although the total gas injection into the pipeline is the same in Cases 1 and 2 (for the same value of $C$), the enhancement of the sensitivity to stochastic gas loads in Case 2 is stronger and more focused for two reasons.  First, the total (aggregated) compression on the system is larger in Case 2.  This can be seen from the curves for compression $\gamma(x)$ in Fig.~\ref{fig:Mass_flux}. Second, the stationary gas loads are more concentrated near $x_*$ resulting in more compression located near $x_*$.  After normalization by $Y$, $Z(x)/Y$ displays a sharper peak.  If the gas load were entirely concentrated at $x_*$, the mass flux and compression would be uniform along the pipeline (on either side of $x_*$), and the peak in $Z(x)/Y$ would be even sharper.

The result that the pipeline shows the highest sensitivity to fluctuations near $x_*$ is not a coincidence. The mass flux in the pipeline exhibits a reversal at this point and the compression changes sign.  It is at the flow reversal that the integral in Eq.~(\ref{eq:mean_square_nondim}) of the Appendix is the largest.  Therefore, in pipelines where the direction of the stationary mass flux is primarily in one direction over long distances, the resulting compression will cause the points of flow reversal to be the most sensitive to pressure fluctuation from stochastic gas draws.  This qualitative result begins to suggest the possibility of fluctuation-aware control algorithms that adjust either the mean gas pressure or the spatial distribution of compression to limit the probability of the gas pressure violating either upper or lower pressure limits.

\begin{figure}
\begin{center}
\includegraphics[width = 3.5 in]{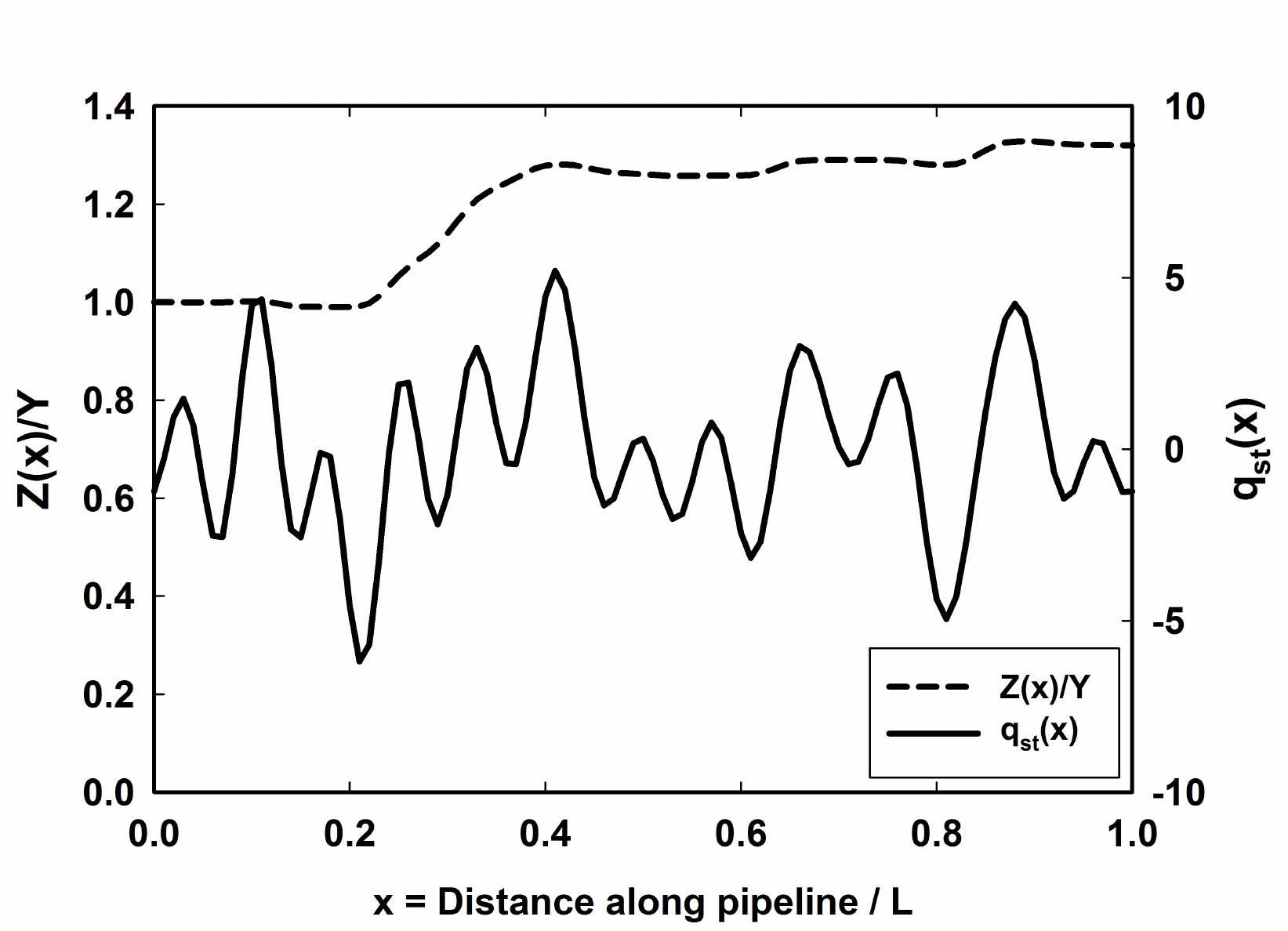}
\end{center}
\caption{Disordered stationary gas injections and loads ($q^{(\mbox{st})}$, solid line) and the fluctuation sensitivity parameter ($Z(x)/Y$, dashed line) versus $x$ for Case 3. The total gas injection into the pipeline is equivalent to $C\approx 14.9$ in Case 1 in Fig.~\ref{fig:ZoverY_linear} or Case 2 in Fig.~\ref{fig:ZoverY_sqrt}.  The disorder of $q^{(\mbox{st})}(x)$ results in many flow reversals that suppress $Z(x)/Y$ suggesting that this configuration is more robust to gas load fluctuations than a pipeline with more unidirectional flow. }
\label{fig:Random}
\end{figure}

Gas injections at the ends of the pipeline do not always dominate the flow in a pipeline. Such a situation may occur near the beginning of a major pipeline where there are many sources of gas injections interspersed with many gas customers. The flow in the pipeline may alternate many times before a significant unidirectional flow builds up.  This situation often occurs in the Williams Transco interstate pipeline near its beginning in Texas\cite{Transco}. This situation may also arise in smaller intrastate pipelines where many smaller, spatially distributed sources are injecting into a pipeline that is serving many different customers.  Case 3 models these configurations by distributing both gas loads and injections along the pipeline with zero injection or load at the ends, i.e. $\phi^{(\mbox{st})}(0)=\phi^{(\mbox{st})}(L)=0$. Fig.~\ref{fig:Random} (solid line) displays a realization of spatially disordered stationary loads and injections $q^{(\mbox{st})}(x)$ that corresponds to a total gas flow equivalent to $C\approx 14.9$ in Case 1 or 2 from above.  Although the total gas injection is similar, the frequent flow reversals limit and the build up of the integral in Eq.~(\ref{eq:mean_square_nondim}) reduces the values of $Z(x)/Y$ in Fig.~\ref{fig:Random} (dashed line) as compared to the Cases 1 and 2 where the flow is more spatially uniform.  The spatial disorder of $q^{(\mbox{st})}(x)$ results in a system that is more robust to fluctuations of gas loads.

\section{Perspectives}
\label{sec:conclusions}

We have developed a dynamical model of natural gas pipelines that incorporates the effect of fluctuating gas injections and loads on the pressure at all points along the pipeline.  The model divides the injections and loads into a stationary component and a fluctuating in time component.  Compressors along the pipeline are adjusted so that the solution for the stationary gas pressure is spatially uniform. An asymptotic solution for the fluctuating pressure factorizes into a product of two terms. The first term depends on the profile of the stationary injection/consumption along the paper and is related to the compression deployed in the stationary solution. Surprisingly, this term does not depend on the gas load fluctuations. The second term grows diffusively in time as given by a spatiotemporal integral of the zero-mean gas load fluctuations. Results for exemplary cases show that the sensitivity of pressure fluctuations to gas load fluctuations is peaked at and around locations of stationary mass flux reversals.  The results suggest the development of a risk-aware gas compressor control that limits the probability of the gas pressure exceeding upper engineering limits or lower contract delivery limits. Pipelines with spatially-disordered injections and loads show less sensitivity to gas load fluctuations.

There are many areas for future work including:
\begin{itemize}
\item The current formulation should be converted to a node-edge network model more amenable to the simulation of real gas networks with compression concentrated at gas compressor stations.
\item Discrete compressor stations will force the relaxation of our assumption of spatially uniform pressure.
\item The solution for the stationary compression should be converted to an optimization for gas pipeline operations (e.g. for minimum cost of compression, maximum throughput, etc) while limiting the probability of violating an upper or lower gas pressure limit.
\end{itemize}

%\begin{acknowledgments}
The authors acknowledge multiple discussions with R. Bent and S. Misra, and support of the Advanced Grid Modeling Program in the Office of Electricity in the U.S. Department of Energy.

\appendix

\section{Description of Appendixes}

The Supplementary Information contained in the Appendixes describes in greater detail the physical models of gas flow and the approximations used to develop the models discussed in the main text.

Methods used to derive main results of the paper are detailed in Appendix \ref{sec:methods}, consisting of two Subsections %\ref{subsec:LinStoch,subsec:prob_analysis}
devoted to discussion of the linearized one-dimensional model of stochastic gas dynamics and following analysis of the probabilistic measure of risk, respectively.

The remaining Appendixes provide discussions of more general modeling needed to support the paper's conclusions. Appendix \ref{sec:pipe} describes basic hydrodynamic equations for a single pipe and discusses the slow transient approximation used in the main text. Appendix \ref{sec:DGF} generalizes the single pipe Dynamic Gas Flow (DGF) description to the case of non-steady gas flows over a meshed network. Appendix \ref{sec:steady} briefly discusses steady Gas Flow (GF) solutions of the DGF model and puts them in the context of the Optimum Gas Flow (OGF) problem used to determine gas compressor operation. In Appendix \ref{sec:pert}, the DGF system is linearized around a steady solution and the general solution of the linear dynamic problem over the network is constructed. The solution is split into homogeneous (zero mode) and inhomogeneous parts and it is argued that the inhomogeneous part of the linearized DGF becomes asymptotically small in the regime of interest. Here we also add a Subsection briefly discussing the inhomogeneous correction, for the general case and then also for the special model of a long pipeline with distributed compression discussed in the main text.

\section{Methods}
\label{sec:methods}

\subsection{Linearized Model of Stochastic Gas Dynamics}
\label{subsec:LinStoch}

The stationary solution described above applies to gas pipelines with well-behaved gas loads. Under these conditions, the pressure $p_0$ does not vary and the pipeline operations are very secure and reliable.  Stochastic gas loads that arise from gas generation compensating fluctuating renewable generation change this picture.
Fluctuating gas loads are added to the stationary solution
\begin{eqnarray} \label{eq:noise}
q^{(\mbox{st})}(x)\to q(t;x)=q^{(\mbox{st})}(x)+\xi(t;x),
 \end{eqnarray}
where $\xi(t;x)$ is zero mean ($\langle\xi\rangle=0$) so that each load, although stochastic, is restricted to consume its scheduled amount $q^{(\mbox{st})}(x)$ over the intra-day market subperiod.  The stochastic component of the gas load $\xi(t;x)$ is expected to include spatiotemporal correlations typical of renewable generation, e.g. $\xi(t;x)$ for wind generation is expected to be correlated on the time scale of tens of minutes to hours over lengths from tens to hundreds of miles.

The effect of the stochastic gas loads is analyzed by linearizing the hydrodynamic model in Eqs.~(\ref{eq:mass_con},\ref{eq:NS}).
(Linearization of the basic non-stationary gas flow equations was already discussed in the literature, however only in the context of simplifying numerical evaluations of the underlying partial differential equations (e.g. \cite{10BS} and references therein). Here, we carry it two steps further--we derive analytical relations and then to analyze effects of stochastic fluctuations and spatial disorder in gas loads. Another recent analytical approach retains the basic nonlinearity but assumes adiabaticity, i.e. very slow changes in the gas loads \cite{11HMS}. Although promising computationally, this approach fails to account for fast, but not necessarily large, fluctuations in the gas draws originating from the electric grid-natural gas pipeline interaction.) Expanding these equations to first order in the fluctuations yields
 \begin{eqnarray}
  c_s^{-2}\partial_t \delta p+\partial_x\delta\phi&=&-\xi,  \label{eq:mass_con_1}\\
 \partial_x \delta p+ \frac{\beta}{d} \frac{\phi^{(\mbox{st})}\delta\phi}{p_0} -\frac{\beta}{d} \frac{(\phi^{(\mbox{st})})^2}{p_0^2}\delta p &=& 0,  \label{eq:NS_1}
 \end{eqnarray}
where $\delta p$ and $\delta \phi$ are the fluctuating pipeline pressures and mass flows, respectively.  Although the gas loads fluctuate, the gas injections $\phi_0$ and $\phi_L$ remain at their stationary values imposing conditions on the fluctuating mass flows
\begin{equation} \label{eq:balance_1}
 \int_0^L dx \ \partial_x \delta \phi=0.
 \end{equation}

The structure of Eqs.~(\ref{eq:mass_con_1}-\ref{eq:balance_1}) provides some guidance regarding the types of solutions expected. Differentiating Eq.~(\ref{eq:NS_1}) with respect to $x$ (and temporarily assuming a uniform $\phi^{(\mbox{st})}$) enables the elimination of $\delta \phi$ via Eq.~(\ref{eq:mass_con_1}).  The resulting PDE in $\delta p$ has the structure of a nonlinear diffusion equation that is driven by exogenous perturbations $\xi(t;x)$.  Because $\langle\xi\rangle=0$ over the intra-day periods, it is tempting to drop all time derivatives in Eq.~(\ref{eq:balance_1}) and compute time-independent mean square fluctuations of $\delta p $, however, this approach is incomplete. Spatiotemporal correlations in $\xi(t;x)$ occurring on time scales shorter than the intra-day period will result in gas draw fluctuations that create shorter-term imbalance with net flow of gas into or out of the pipeline.  Eq.~(\ref{eq:balance_1}) shows that these non-zero net fluctuations are not allowed to leak out either end of the pipeline because $\phi_0$ and $\phi_L$ are held fixed.  The only way for the system to accommodate these short-term correlated fluctuations is through a ``zero mode'' where the average pressure in the entire pipeline raises or falls along with the fluctuating injections.  This zero mode and its effects on the pipeline pressure fluctuations are the emphasis of the remainder of this manuscript.

\subsection{Probabilistic Measure of Risk: Analysis}
\label{subsec:prob_analysis}

To represent the slow drifts of pipeline pressure discussed above, we suggest a solution to Eqs.~(\ref{eq:mass_con_1},\ref{eq:NS_1}) of the form
 \begin{eqnarray} \label{eq:solution_form}
 \delta p(t;x) = a(t) Z(x) + b(t;x),
 \end{eqnarray}
where the two components of the solution respond to the different characteristics of the fluctuations $\xi(t;x)$.  The first component $a(t) Z(x)$ is the zero mode where $Z(x)$ is a slowly varying function of $x$ that captures the spatial distribution of gas stored in the pipeline as pressure rise or fall driven by the correlated component of the fluctuating gas draws. The time dependence of these global pressure swings are captured by $a(t)$.  In contrast, $b(t;x)$ varies more rapidly in space and responds to the uncorrelated component of the fluctuations of $\xi(t;x)$ that occur on finer spatial and time scales.  The larger spatial derivatives of $b(t;x)$ result in relatively rapid diffusion of pressure (and gas) which limits the impact of $b(t;x)$ on pressure fluctuations.

Substituting our proposed solution (\ref{eq:solution_form}) into Eqs.~(\ref{eq:mass_con_1},\ref{eq:NS_1}) yields an equation for the zero mode
\begin{equation}\label{eq:Z_eq}
\partial_x Z-\frac{\beta}{d} \frac{\phi^{(\mbox{st})}(x^\prime)|\phi^{(\mbox{st})}(x^\prime)|}{p_0^2}Z = 0,
\end{equation}
which has a solution
\begin{equation}\label{eq:Z_solution}
 Z(x) =\exp \left[ \int_0^x dx^\prime\ \frac{\beta\phi^{(\mbox{st})}|\phi^{(\mbox{st})}|}{d\,p_0^2}\right]=\exp \left[ \int_0^x 2 \gamma(x^\prime) dx^\prime \right] .
\end{equation}
The solution for $Z(x)$ does not depend on the form of the fluctuations $\xi$.  Rather, it depends on the stationary solution $\phi^{(\mbox{st})}(x)$, or equivalently on deployed gas compression in the stationary solution.  The same substitution also yields an expression for $a(t)$ that does depend on the gas load fluctuations:
\begin{equation}\label{eq:a_eq}
c_s^{-2}Z\partial_t a+\delta_x\delta\phi=-\xi.
\end{equation}
Since $a(t)$ is independent of $x$, Eq.~(\ref{eq:a_eq}) can be integrated over the length of the pipeline to yield an explicit expression for $a(t)$:
\begin{equation} \label{eq:a_solution}
a=-\frac{c_s^2}{L Y} \int_0^t dt'\int_0^L dx\ \xi(t',x),\quad Y=\int_0^L dx\ Z/L,
 \end{equation}
where Eq.~(\ref{eq:balance_1}) has been used to eliminate the $\delta \phi$ term.

The physical interpretation of the zero mode $a(t)Z(x)$ now becomes clear.  The double integral in Eq.~(\ref{eq:a_solution}) filters out the uncorrelated components of $\xi$ showing that the time dependence of the zero mode $a(t)$ only responds to the fluctuations of $\xi$ that are correlated in space (over the entire length of the pipeline) and in time (since the beginning of the intra-day market period). A discussion of the solution component $b(t;x)$ is given below in Section \ref{subsec:inhomogeneous}. % in Section \ref{sec:inhomogeneous} of the SM.

The zero mode $a(t)Z(x)$ will dominate the contribution to $\delta p(t)$ at times longer than the correlation time $\tau$ of $\xi$ where $\tau$ is expected to be in the range of tens of minutes to hours for fluctuating gas loads creating by gas-fired electric generators balancing intermittent wind generation. For $t \gg \tau$, we may safely drop the $b$ in favor of $a(t)Z(x)$ and estimate the pressure variation covariance as
 \begin{eqnarray}
 && \langle (\delta p(x))^2 \rangle=\label{eq:mean_square}\\
 && \frac{c_s^4\tau t}{L^2} \left(\frac{Z(x)}{Y}\right)^2\iint_0^L dx_1\ dx_2 \langle \xi(t,x_1)\xi(t,x_2) \rangle,
 \nonumber
 \end{eqnarray}
where we have also assumed statistical stationarity of $\xi(t;x)$ over time.

Eq.~(\ref{eq:mean_square}) can be rearranged slightly to reveal a physical interpretation, as shown in Fig.~(6) of the main text.
%i.e.
%\begin{eqnarray}
%\frac{\langle (\delta p(x))^2 \rangle}{p_0^2} =\left(\frac{\overline{q}^{(\mbox{st})}c_s^2 \tau}{p_0^2}\right)^2 \frac{t}{\tau}\left(\frac{Z(x)}{Y}\right)^2 \frac{1}{L^2}\iint_0^L dx_1\ dx_2 \frac{\langle \xi(t,x_1)\xi(t,x_2)\rangle}{(\overline{q}^{(\mbox{st})})^2}.
%\label{eq:mean_square_nondim}
%\end{eqnarray}
The first term on the right hand side of Eq.~(6) of the main text
%(\ref{eq:mean_square_nondim})
is the square of the fractional pressure decline if the entire pipeline was subject to the spatially averaged gas load $\overline{q}^{(\mbox{st})}$ without any compensating injections for one correlation time $\tau$ of the gas load fluctuations.  This first term is multiplied by the number of correlation times ($t/\tau$) since the intra-day period began.  The third term provides the only $x$ dependence and describes the sensitivity of different locations in the pipeline to pressure fluctuations.  This dependence comes entirely through $Z(x)$ which (see Eq.~(\ref{eq:Z_solution})) depends only on the compression deployed in the stationary solution.  The dependence on $Z(x)$ demonstrates that a highly stressed pipeline, i.e. one with a large $\int \gamma(x^\prime) dx^\prime$, is more susceptible to pressure fluctuations driven by stochastic gas loads, and $Z(x)$ shows which pipeline locations are most susceptible. The final term on the the right hand side of Eq.~(6) of the main text measures the spatial average of the correlated fluctuations in the gas loads normalized by the average stationary gas load.

The right hand side of expression for $a(t)$ in Eq.~(\ref{eq:a_solution}) is a time integral over a stochastic process, and per the law of large numbers, $a(t;x)$ and $\delta p(t;x)$ are expected to be asymptotically Gaussian when the integration time is longer than the correlation time of $\xi$. In this limit, the estimate of the pressure fluctuation covariance in Eq.~(\ref{eq:mean_square}) or Eq.~(6) of the main text also predicts the tails of the distribution over $\delta p$, thus allowing the estimation of the probability of relatively rare events of high or low pressure fluctuations (under the condition that the fluctuations are still within the linear approximation used here).  Eq.~(\ref{eq:mean_square}) becomes a probabilistic measure of risk to reliability of natural gas pipeline operations and a route to modeling the risk that cascades to the interdependent infrastructures such as electric power systems.

\section{Gas Flow Equations: Individual Pipe}
\label{sec:pipe}

Following \cite{13MFBBCP}, we consider the flow of a compressible gas in a single length of pipe. Major transmission pipelines are typically 16-48 inches in diameter and operate at high pressures and mass flows, e.g. $200$ to $1500$ pounds per square inch (psi) and moving millions of cubic feet of gas per day \cite{Ref_Crane1982,Ref_Mokhatab2006}. Under these highly turbulent conditions, the pressure drop and energy loss due to shear is well represented by a nearly constant phenomenological friction factor $f$.  The resulting gas flow model is a partial differential equation (PDE) with one spatial dimension $x$ (along the pipe axis) and one time dimension \cite{osiadacz1987simulation,87TT,05Sar}:
\begin{eqnarray}
&& \partial_t\rho+\partial_x (u\rho)=0,\label{density_eq}\\
&& \partial_t (\rho u)\!+\!\partial_x (\rho u^2)\!+\!\partial_x p\!=-\frac{\rho u |u|}{2d} f\!-\!\rho g \sin\alpha,\label{momenta_eq}\\
&& p=\rho Z R T.\label{thermodynamic_eq}
\label{state_eq}
\end{eqnarray}
Here, $u,p,\rho$ are velocity, pressure, and density at position $x$; $Z$ is the gas compressibility factor; $T$ is the temperature, $R$ is the gas constant, and $d$ is the diameter of the pipe.

Eqs.~(\ref{density_eq},\ref{momenta_eq},\ref{state_eq}) represent mass conservation, momentum balance and the ideal gas thermodynamic relation, respectively. The first term on the rhs of Eq.~(\ref{momenta_eq}) represents the friction losses in the pipe. The second term on the rhs of Eq.~(\ref{momenta_eq}) accounts for the gain or loss of momentum due to gravity $g$ if the pipe is tilted by angle $\alpha$.  The frictional losses typically dominate the gravitational term, which is typically dropped.  Similarly, the gas inertia term ($\partial_t(\rho u)$ is also typically small compared to the frictional losses (as the flow velocity is significantly smaller than sound velocity) and is dropped. For simplicity of presentation, we have also assumed that the temperature does not change significantly along the pipe.

Taking into account these assumptions, Eqs.~(\ref{density_eq},\ref{momenta_eq},\ref{thermodynamic_eq}) are rewritten in terms of the pressure $p$ and the mass flux $\phi=u\rho$:
\begin{eqnarray}
&& c_s^{-2}\partial_t p + \partial_x\phi=0,\label{density_eq1}\\
&& \partial_x p+ \frac{\beta}{2d}\frac{\phi|\phi|}{p}=0,\label{momenta_eq1}
\end{eqnarray}
where $c_s\equiv \sqrt{ZRT}$ is the sound velocity and $\beta\equiv f Z R T$ are both considered constant. To resolve the dynamic problem for $t\in[0,\tau]$ over $x\in[0,L]$ we also need to supply Eqs.~(\ref{density_eq1},\ref{momenta_eq1}) with initial and boundary conditions for flows,
\begin{eqnarray}
&& t=0,\quad \forall x\in[0,L]:\quad \phi(0;x)=\phi_0(x),\label{phi-0}\\
&& \forall t:\quad \phi(t;0)=q^{(\mbox{\small in})}(t),\quad \phi(t;L)=q^{(\mbox{\small out})}(t),
\label{phi-in-out}
\end{eqnarray}
which are consistent, i.e. $\phi_0(0)=q^{(\mbox{\small in})}(0)$ and $\phi_0(L)=q^{(\mbox{\small out})}(0)$,
in addition fixing pressure initially at an end of the pipe, e.g. $p(0;0)=p_0$.

%\begin{eqnarray}
%&& t=0,\ \forall x\in[0,L]:\quad p(0;x)=p^{(in)}(x), \label{p-initial}\\
%&& \forall t,\ x=0:\quad p(t;0)=p^{(sl)}(t), \label{p-sl}
%\end{eqnarray}
%defining the initial pressure (and therefore mass flow) throughout the pipe and the controlled pressure $p^{(sl)}(t)$) at the beginning of the pipe, respectively.

\section{Dynamic Gas Flow (DGF) over Network}
\label{sec:DGF}

%\begin{multicols}{1}
\begin{figure*}%[h]
\centering
\includegraphics[width=0.85\textwidth]{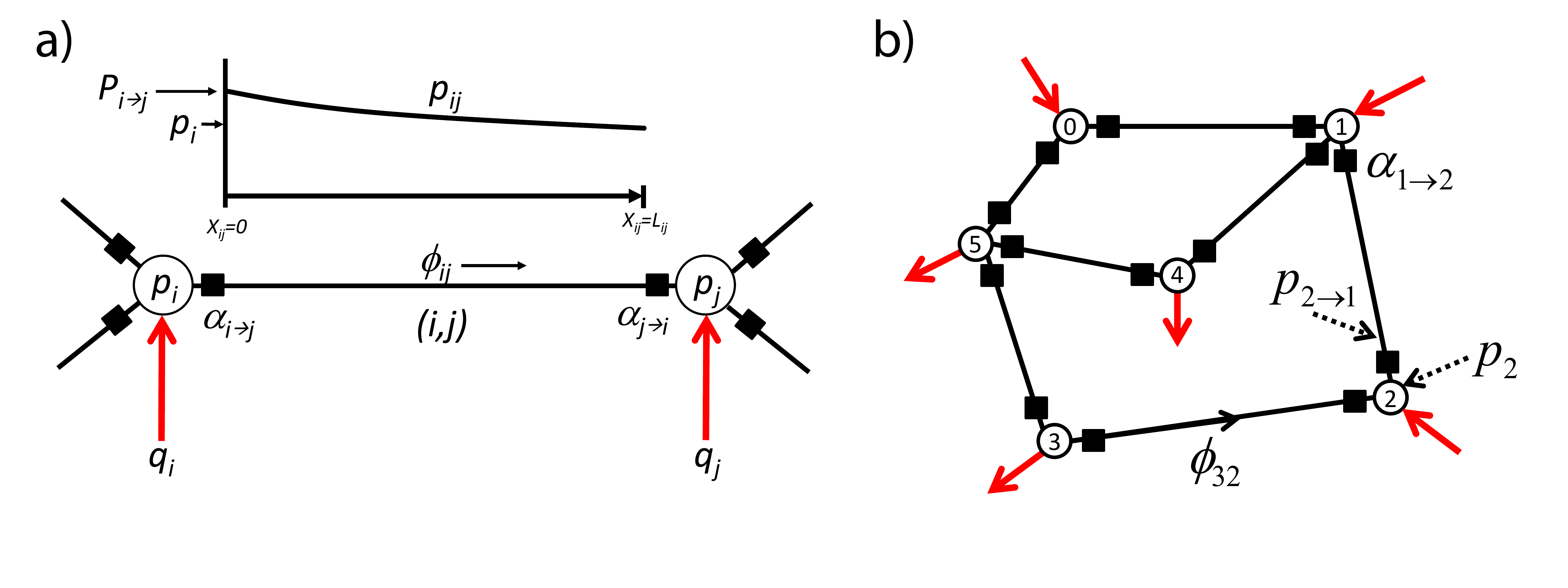}
\caption{Schematic illustration of the network-structure notations. a) Schematic illustration of a single edge $(i,j)$ of a network. Nodes at either end are indicated by open circles and labeled by their nodal pressure $p_i$ and $p_j$. Compressors are indicated with filled squares. Mass flow $\phi_{ij}$ is directed from $i$ to $j$ and injections $q_i$ and $q_j$ contribute to this flow. Nodal pressure $p_i$ is modified by the compression ratio $\alpha_{i\to j}$ yielding $p_{ij}(x_{ij}=0)$.  The pressure falls along $\{i,j\}$ reaching $p_{ij}(x_{ij}=L_{ij})$.  If compressor $\alpha_{j\to i}$ is not present, then $p_{ij}(x_{ij}=L_{ij})=p_j$.  b) Schematic of many edges connected in a meshed network.  Nodes are indexed by $i=0,1,\cdots$, where node $0$ is typically reserved for the swing bus -- the node where pressure is maintained constant throughout the dynamics. Compressors and injections and edge mass flows are the same as in a).
}
\label{fig:scheme}
\end{figure*}
%\end{multicols}

The single pipe setting in Eqs.~\ref{density_eq1} and \ref{momenta_eq1} is generalized to a gas network represented by a graph ${\cal G}=({\cal V},{\cal E})$ with a set of vertexes ${\cal V}$ and set of edges ${\cal E}$, where the edges will be considered directed or undirected, depending on the context. See Fig.~\ref{fig:scheme} for a schematic illustration. We will adopt $(i,j)$ and $\{i,j\}$ notations for directed and undirected edges, respectively. Each vertex, $i\in{\cal V}$ represents a node with a gas injection/consumption rate $q_i$ (mass per unit time). Each edge $(i,j)\in{\cal E}$ is a single pipe with mass flow $\phi_{ij}$. The flow along each edge is described by a set of PDEs:
\begin{eqnarray}
&&\forall t\in [0,\tau],\quad \forall \{i,j\}\in{\cal E},\quad \forall x\in[0;L_{ij}]:\nonumber\\
&&c_s^{-2}\partial_t p_{ij}(t,x)+\partial_x \phi_{ij}(t,x)=0,\label{density_eq1}\\
&& \partial_x p_{ij}(t,x)+\frac{\beta}{2d} \frac{\phi_{ij}(t,x) |\phi_{ij}(t,x)|}{p_{ij}(t,x)}=0,\label{momenta_eq1}
\end{eqnarray}
where $p_{ij}(t,x)$ and $\phi_{ij}(t,x)$ are the pressure and mass flow, respectively, at time $t$ and position $x$ along edge $(i,j)$ of length $L_{ij}$. Here, $p_{ij}=p_{ji}$, $\phi_{ij}=-\phi_{ji}$, and $L_{ij}=L_{ji}$.  See Fig.~\ref{fig:scheme}a for a schematic description of the variables.

The flow of gas create a pressure drop. To compensate, the pressure is boosted at compressor stations potentially located at both ends of each edge $\{i,j\}$.  $\alpha_{i\to j}$ is the compression ratio of the station adjacent to node $i$ while $\alpha_{j\to i}$ is the compression ratio adjacent to node $j$. We choose to place compressors at the two ends of every line/pipe for generality, which also simplifies the notations in the following discussion. In reality there will be only none or one compressor on any particular edge of the graph. Note also that $\alpha_{i\to j}$ may be larger or smaller than unity,  thus representing compression or decompression.  If only compression is allowed, then $\alpha_{i\to j}\geq 1$. The relationships between the pressures in Fig.~\ref{fig:scheme} are
\begin{eqnarray}
&&\forall t\in [0,\tau],\quad \forall (i,j)\in{\cal E}:\quad p_{ij}(t,0)=p_{i\to j}(t), \label{compressor} \\
&& p_{ij}(t,L_{ij})=p_{j\to i}(t),\ p_{i\to j}=p_i \alpha_{i\to j},\ p_{j\to i}=p_j \alpha_{j\to i}, \nonumber
\end{eqnarray}
where $p_i$ and $p_{i\to j}$ are the pressures at node $i$ and just past the compressor on edge $(i,j)$ adjacent to node $i$ and the last part of Eq.~(\ref{compressor}) is added for clarity.  If there is no compressor installed at the beginning of the edge $(i,j)$ or if the compressor is inactive, $\alpha_{i\to j}=1$. In the current operational paradigm, compression rates are not changed very frequently, however, we anticipate changes and allow the $\alpha_{i\to j}$ to depend on time.

Eqs.~(\ref{momenta_eq1},\ref{compressor}) are complemented with mass conservation at all nodes of the network:
\begin{equation}
\forall t\in [0,\tau],\quad \forall i\in{\cal V}: \sum_{j:(i,j)\in{\cal E}}\phi_{ij}(t,0)=q_i(t).
\label{flow_conserve}
\end{equation}
When the gas injections $q(t)=(q_i(t)|i\in{\cal V})$ for are given for $t\in[0,\tau]$, nodal conditions (\ref{flow_conserve}) generalize the single-pipe boundary conditions in (\ref{phi-in-out}) to a pipe network. Eqs.~(\ref{density_eq1},\ref{momenta_eq1},\ref{compressor},\ref{flow_conserve}) constitute a complete set of equations describing the Dynamic Gas Flow (DGF) problem if they are supplemented with compression ratios, i.e. $\alpha=(\alpha_{i\to j}|(i,j)\in{\cal E})$, initial conditions on the flows
\begin{eqnarray}
&& t=0,\ \forall \{i,j\}\in{\cal E},\ \forall x_{ij}\in[0,L_{ij}]:\nonumber\\
&& \phi_{ij}(0;x_{ij})=\phi^{(in)}_{ij}(x_{ij}), \label{phi-initial-netw}
\end{eqnarray}
and pressure at one arbitrarily chosen slack node, $p_{i=0}(0)=p_0$.

%\begin{eqnarray}
%&& t=0,\ \forall i\in{\cal V},\ \forall x_{ij}\in[0,L_{ij}]:\quad p_{ij}(0;x_{ij})=p^{(in)}_{ij}(x_{ij}), \label{p-initial-netw}\\
%&& \forall t,\ i=0:\quad p_0(t)=p^{(\mbox{sl})}(t), \label{p-sl-netw}
%\end{eqnarray}
%where $i=0$ is a slack (swing) node where the pressure is controlled. Naturally, one requires,
%$\forall j\mbox{ s.t.} (0,j)\in{\cal E}:\ p^{(in)}_{0j}(0)=p^{(\mbox{sl})}(0)$, for consistency of Eqs.~(\ref{p-initial-netw}) with Eq.~(\ref{p-sl-netw}).

\section{Stationary Gas Flow and Optimum Gas Flow}
\label{sec:steady}

The stationary/steady version of the DGF problem is the Gas Flow (GF) problem. In the GF problem, all input parameters (consumptions/injections, compression ratios and the pressure at the slack bus) are constant in time, and the total injection and consumption are balanced
\begin{equation}
\sum_{i\in{\cal V}} q_i^{(\mbox{st})}=0. \label{ballance}
\end{equation}
The steady solution of Eq.~(\ref{density_eq1}) is uniform mass flow along each pipe in the network, $\forall \{i,j\}:\quad \phi_{i\to j}=\mbox{const}$. Substituting this result into Eq.~(\ref{momenta_eq1}) and taking straightforward spatial integration yields algebraic relations between flow through and pressures at both ends of every pipe in the network
\begin{eqnarray}
&& \hspace{-0.5cm}\forall (i,j)\in{\cal E}:\quad p_{i\to j}^{(\mbox{st})}=p_i^{(\mbox{st})} \alpha_{i\to j}; \nonumber\\ &&
(p_{ij}^{(\mbox{st})}(x))^2=(p_{i\to j}^{(\mbox{st})})^2-\frac{\beta x}{d} \phi_{ij}^{(\mbox{st})}|\phi_{ij}^{(\mbox{st})}|.
\label{p-phi-steady}
\end{eqnarray}
The GF problem has a unique solution provided the compression ratios are known.
%The GF problem is the basis for many approaches to solving the OGF problem.
%These equations should be supplemented by solutions of the steady versions of Eq.~(\ref{flow_conserve}). The steady GF problem requires one less boundary condition, and $p_0=p^{(\mbox{sl})}$ is sufficient. %{\color{red} Notation not updated?  Should the superscript be st? }
%Note, that according to \cite{77Nau,12BNV} the GF problem can be stated as a computationally feasible (even on graph with loops) convex optimization problem.

Compression ratios $\alpha$ are time-independent in the steady GF setting. The configuration of $\alpha$ over the network is typically decided using a combination of economic and operational factors. The model selected in the main text corresponds to a simple greedy approach, i.e. maintain constant pressure throughout the network for flows corresponding to the forecasted comsumptions/injections. This model roughly replicates the behavior of pipeline operators in the U.S. where the energy consumed in the compression of the gas is not a major concern.
%which is popular with grid operators in US, focused primarily on maintaining reasonable operational constraints for the forecasted/steady solution --- constancy of pressure throughout the system.
More sophisticated compression dispatch options, in particular minimization of the total work spent on compression subject to maintaining pressure within acceptable limits, have been extensively discussed in the literature, e.g. \cite{68WL,00WRBS,10Bor,13MFBBCP} and references therein.

\section{Perturbative solution of the DGF problem}
\label{sec:pert}

We generalize discussion in the main text by introducing stochastic gas loads (due, e.g., to natural gas-fired generators) from a line to a network, such that $q(t)=q^{(\mbox{st})}+\xi(t)$ where components of $\xi(t)=(\xi_i(t)|i\in{\cal V})$ are time varying but relatively small in comparison with $q^{(\mbox{st})}$. We look for a linearized solution of the DGF problem of the form $p(t)=p^{(\mbox{st})}+\delta p(t)$ and $\phi(t)=\phi^{(\mbox{st})}+\delta \phi(t)$,  where the respective corrections are small, i.e. $|\delta p(t)|\ll p^{(\mbox{st})}$ and $|\delta \phi(t)|\ll \phi^{(\mbox{st})}$. The linearized versions of Eqs.~(\ref{density_eq1},\ref{momenta_eq1},\ref{compressor},\ref{flow_conserve}) become
\begin{eqnarray}
&&\hspace{-0.5cm}\forall t\in [0,\tau],\quad \forall \{i,j\}\in{\cal E},\quad \forall x\in[0;L_{ij}]:\nonumber\\ && c_s^{-2}\partial_t \delta p_{ij}+\partial_x \delta \phi_{ij}=0,\label{density_delta}\\
&& \partial_x \delta p_{ij}+\frac{\beta}{2d} \Biggl(\frac{\delta\phi_{ij} |\phi_{ij}^{(\mbox{st})}|}{p_{ij}^{(\mbox{st})}}\nonumber\\ && +
\frac{\phi_{ij}^{(\mbox{st})} |\delta\phi_{ij}|}{p_{ij}^{(\mbox{st})}}-\frac{\delta p_{ij}\phi_{ij}^{(\mbox{st})} |\phi_{ij}^{(\mbox{st})}|}{(p_{ij}^{(\mbox{st})})^2}\Biggr)=0,\label{momenta_delta}\\
&&\hspace{-0.5cm}\forall t\in [0,\tau],\quad \forall (i,j)\in{\cal E}:\quad \delta p_{i\to j}=\delta p_i \alpha_{i\to j} , \label{compressor_delta}\\
&& \delta p_{ij}(t,0)=\delta p_{i\to j}(t),\quad \delta p_{ij}(t,L_{ij})= \delta p_{j\to i}(t),\label{conditions_delta_p}\\
&&\hspace{-0.5cm}\forall t\in [0,\tau],\quad \forall i\in{\cal V}:\quad
\sum_{j:(i,j)\in{\cal E}}\delta \phi_{ij}(t,0)=\xi_i(t).
\label{conditions_delta_phi}
\end{eqnarray}

The remainder of the Subsection is devoted to finding an asymptotic solution of Eqs.~(\ref{density_delta},\ref{momenta_delta},\ref{compressor_delta},\ref{conditions_delta_p},\ref{conditions_delta_phi}).  Here, asymptotic implies finding solutions for time $\tau$ longer than the correlation time of the fluctuation consumption $\xi$. We seek solutions that eliminate the complexity of the PDE of Eqs.~(\ref{density_delta}, \ref{momenta_delta},\ref{compressor_delta},\ref{conditions_delta_p},\ref{conditions_delta_phi}) and that connect the nodal quantities by algebraic relationships.

Therefore, generalizing  the solution proposed in the main text (see Eq. (10)), we look for a solution of Eqs.~(\ref{density_delta},\ref{momenta_delta}) of the form
\begin{eqnarray}
&& \delta p_{ij}=a_{ij}(t)Z_{ij}(x)+b_{ij}(t,x),\label{delta-p}
\end{eqnarray}
where $a_{ij}(t)$ only depends on time. Here in Eq.~(\ref{delta-p}) $Z_{ij}(x)$ solves the following linear homogeneous equation
\begin{equation}
\partial_x Z_{ij}-\frac{\beta}{d}\frac{\phi_{ij}^{(\mbox{st})}|\phi_{ij}^{(\mbox{st})}|}{p_{ij}^{(\mbox{st})}}Z_{ij}=0,
\label{Z}
\end{equation}
where $Z_{ij}(x)$ counts $x$ from node $i$, i.e. reversing the direction of counting one gets, $Z_{ij}(L_{ij})=Z_{ji}(0)$.

Assuming that $\tau$ is sufficiently large, we conjecture (which will be verified after the global asymptotic solution is found) that the major contribution to $\delta p_{ij}$ in Eq.~(\ref{delta-p}) originates from the first ``zero-mode" term $a_{ij}(t)Z_{ij}(x)$ that (as will be seen below) grows in time compared to the second term that does not.

To find the leading (zero mode) term we proceed as follows. The integration of Eq.~(\ref{Z}) over the spatial dependence of the stationary profile (\ref{p-phi-steady}),  yields
\begin{eqnarray}
Z_{ij}(x)=\frac{p_{i\to j}^{(\mbox{st})}+p_{j\to i}^{(\mbox{st})}}{2p_{ij}^{(\mbox{st})}(x)},
\label{Z2}
\end{eqnarray}
where the normalization constant is chosen to guarantee, $\int_0^L Z_{ij}(x) dx/L=1$.
We solve for the time-dependent factor $a_{ij}(t)$ by substituting $\delta p_{ij}$ with $a_{ij}(t)Z_{ij}(x)$ into Eq.~(\ref{density_delta}) and integrate the result over the entire spatial extent of the pipe $\{i,j\}$ yielding
\begin{eqnarray}
&& a_{ij}(t)=c_s^2\int_0^tdt'\left(\delta\phi_{ij}(t',0)-\delta\phi_{ij}(t',L)\right).
\label{a-via-delta-phi}
\end{eqnarray}
In the asymptotic limit where $\delta p_{ij}$ is approximated by $a_{ij}(t)Z_{ij}(x)$ for every pipe (graph edge), Eqs.~(\ref{conditions_delta_p}) can only be satisfied if the $a_{ij}(t)$ have the same functional dependence on time, i.e.,
\begin{equation}
\forall \{i,j\}\in{\cal E}:\quad a_{ij}(t)=a(t)c_{ij},\label{a}
\end{equation}
where $c_{ij}=c_{ji}$ is an edge specific constant. To compute the global time-dependent factor $a(t)$ in Eq.~(\ref{a}) we sum over all the nodes of the graph
\begin{equation}
\sum_{i\in{\cal V}}\xi_i=\sum_{\{i,j\}\in{\cal E}}\left(\delta\phi_{ij}(t,0)-\delta\phi_{ij}(t,L_{ij})\right),
\label{sum_delta_phi}
\end{equation}
integrate over time, define
\begin{equation}
\Xi(t)\doteq\int_0^tdt' \sum_{i\in{\cal V}}\xi_i(t'),\label{Xi}
\end{equation}
and finally sum Eq.~(\ref{a}) overall edges:
\begin{equation}
a(t)=\frac{c_s^2 \Xi(t)}{\sum_{\{i,j\}\in{\cal E}}c_{ij}}.
\label{aa}
\end{equation}
Therefore, $\forall t,\ \ \forall \{i,j\}\in{\cal E},\ \ x\in[0,L_{ij}]:$
\begin{equation}
\delta p_{ij}(t,x)\approx \frac{c_s^2\Xi(t)}{\sum_{\{i,j\}\in{\cal E}}c_{ij}} c_{ij}Z_{ij}(x).\label{delta_p_final}
\end{equation}
The unknown edge constants $c_{ij}$ are derived by substituting Eqs.~(\ref{delta_p_final}) into Eqs.~(\ref{compressor_delta}, \ref{a}) yielding
\begin{equation}
\forall i,\ \ \forall j,k\ \ \mbox{s.t.  }(i,j),(i,k)\in{\cal E}:\ \ \frac{c_{ij} Z_{ij}(0)}{\alpha_{i\to j}}=\frac{c_{ik}Z_{ik}(0)}{\alpha_{i\to k}}.
\label{c-relations}
\end{equation}
Eqs.~(\ref{delta_p_final}, \ref{c-relations}, \ref{Z2}) express the complete asymptotic (zero mode) solution of the DGF problem.

Assuming that the random gas load fluctuations $\xi_i(t)$ are zero-mean, temporarily homogeneous, and  relatively short correlated in both time (the correlation time is less than $\tau$) and space (the correlation length is less than the spatial extent of the network), and observing that $\delta p_{ij}$ in Eq.~(\ref{delta_p_final}) is given by a time-integral and spatial-sum of the fluctuations, one concludes that according to the Large Deviation theory, {\em the pressure fluctuations form a Gaussian random process which jitter diffusively in time},  i.e. the Probability Distribution Function (PDF) of $\delta p_{ij}(t,x)$ is
\begin{eqnarray}
&& {\cal P}(\delta p_{ij}(t,x)=\delta)\to \nonumber\\
&&\left(2\pi t D_{ij}(x)\right)^{-1/2}
\exp\left(-\frac{\delta^2}{2 t D_{ij}(x)}\right),
\label{PDF}\\
&&  D_{ij}=\left(\frac{c_s^2 c_{ij}Z_{ij}(x)}{\sum_{\{k,l\}\in{\cal E}}c_{kl}}\right)^2
\Biggl\langle \left(\sum_{n\in{\cal V}}\xi_n(t')\right)^2\Biggr\rangle
,\label{xi-xi}
\end{eqnarray}
where the correlation function on the right-hand-side does not depend on $t'$  due to assumption of the statistical homogeneity of $\xi$.

\subsection{Correction to the asymptotic solution}
\label{subsec:inhomogeneous}

In the general analysis of the preceding Section of this SI, the pressure fluctuations are separated into homogeneous (zero mode) and inhomogeneous (forced) components, according to Eq.~(\ref{delta-p}).  The formal separation in Eq.~(\ref{delta-p}) leads to a differential equation for the inhomogenous solution $b(t;x)$ which, for the general formulation above, is fully defined by Eqs.~(\ref{b-eq},\ref{delta-phi-eq}).

Once the leading, growing in time, contribution to $\delta p_{ij}$ is found,
one verifies that, $b_{ij}(t,x)$, extracted from
\begin{eqnarray}
&& \partial_x b_{ij}+\frac{\beta}{2d} \Biggl(\frac{\delta\phi_{ij} |\phi_{ij}^{(\mbox{st})}|}{p_{ij}^{(\mbox{st})}}
\nonumber\\ && +
\frac{\phi_{ij}^{(\mbox{st})} |\delta\phi_{ij}|}{p_{ij}^{(\mbox{st})}}-\frac{\delta p_{ij}\phi_{ij}^{(\mbox{st})} |\phi_{ij}^{(\mbox{st})}|}{(p_{ij}^{(\mbox{st})})^2}\Biggr)=0.
\label{b-eq}\\
&& c_s^{-2}Z_{ij}\frac{d}{dt}a_{ij}+\partial_x \delta \phi_{ij}=0, \label{delta-phi-eq}
\end{eqnarray}
does not grow with time, and thus it is asymptotically smaller --- consistently with what was conjectured above to derive the leading contribution.

Let us discuss this asymptotic separation of the solution into dominant contribution and correction in more details for the simplified analysis/model of the main text. Repeating the solution separation on the (simplified) continuous-compression model, we find a differential equation for $b(t;x)$ in terms of $\delta \phi$, i.e.
\begin{equation}\label{eq:b_eq}
\partial_x b+ \frac{\beta}{d} \frac{|\phi^{\mbox{st}}|\delta\phi+\phi^{\mbox{st}}|\delta\phi|}{2p_0} -\frac{\beta}{d} \frac{(\phi^{\mbox{st}})^2}{p_0^2}b = 0.
\end{equation}
Fluctuations in $\delta \phi$ drive $b(t;x)$, but unlike for the homogeneous solution, Eq.~(\ref{eq:b_eq}) shows that this effect is entirely local.  Specifically, Eq. (14) of the main text shows that the homogeneous component responds to the global imbalance in gas loads while the response in Eq.~(\ref{eq:b_eq}) is to the local $\delta \phi$.  In addition, the response in Eq.~(\ref{eq:b_eq}) decays in space and does so quickly in areas of high compression for the stationary solution (see Eq. (5) of the main text).  In contrast, the homogeneous solution grows more quickly in areas of high compression (see Eq. (12) of the main text).  These two properties contribute to the dominance of the homogeneous solution over the inhomogeneous solution at longer times.

\bibliographystyle{apsrev4-1}
%\bibliographystyle{aipauth4-1}
%\bibliography{../Bib/GasFlow,../Bib/Russian}
\bibliography{../Bib/GasFlow,../Bib/Russian,../Bib/RefConrado,../Bib/GasFlowSM}

\begin{thebibliography}{19}%
\makeatletter
\providecommand \@ifxundefined [1]{%
 \@ifx{#1\undefined}
}%
\providecommand \@ifnum [1]{%
 \ifnum #1\expandafter \@firstoftwo
 \else \expandafter \@secondoftwo
 \fi
}%
\providecommand \@ifx [1]{%
 \ifx #1\expandafter \@firstoftwo
 \else \expandafter \@secondoftwo
 \fi
}%
\providecommand \natexlab [1]{#1}%
\providecommand \enquote  [1]{``#1''}%
\providecommand \bibnamefont  [1]{#1}%
\providecommand \bibfnamefont [1]{#1}%
\providecommand \citenamefont [1]{#1}%
\providecommand \href@noop [0]{\@secondoftwo}%
\providecommand \href [0]{\begingroup \@sanitize@url \@href}%
\providecommand \@href[1]{\@@startlink{#1}\@@href}%
\providecommand \@@href[1]{\endgroup#1\@@endlink}%
\providecommand \@sanitize@url [0]{\catcode `\\12\catcode `\$12\catcode
  `\&12\catcode `\#12\catcode `\^12\catcode `\_12\catcode `\%12\relax}%
\providecommand \@@startlink[1]{}%
\providecommand \@@endlink[0]{}%
\providecommand \url  [0]{\begingroup\@sanitize@url \@url }%
\providecommand \@url [1]{\endgroup\@href {#1}{\urlprefix }}%
\providecommand \urlprefix  [0]{URL }%
\providecommand \Eprint [0]{\href }%
\providecommand \doibase [0]{http://dx.doi.org/}%
\providecommand \selectlanguage [0]{\@gobble}%
\providecommand \bibinfo  [0]{\@secondoftwo}%
\providecommand \bibfield  [0]{\@secondoftwo}%
\providecommand \translation [1]{[#1]}%
\providecommand \BibitemOpen [0]{}%
\providecommand \bibitemStop [0]{}%
\providecommand \bibitemNoStop [0]{.\EOS\space}%
\providecommand \EOS [0]{\spacefactor3000\relax}%
\providecommand \BibitemShut  [1]{\csname bibitem#1\endcsname}%
\let\auto@bib@innerbib\@empty
%</preamble>
\bibitem [{\citenamefont {Arthur}\ \emph {et~al.}()\citenamefont {Arthur},
  \citenamefont {Langhus},\ and\ \citenamefont {Alleman}}]{fracking}%
  \BibitemOpen
  \bibfield  {author} {\bibinfo {author} {\bibfnamefont {J.}~\bibnamefont
  {Arthur}}, \bibinfo {author} {\bibfnamefont {B.}~\bibnamefont {Langhus}}, \
  and\ \bibinfo {author} {\bibfnamefont {D.}~\bibnamefont {Alleman}},\
  }\href@noop {} {\enquote {\bibinfo {title} {An overview of modern shale gas
  development in the united states},}\ }\BibitemShut {NoStop}%
\bibitem [{ISO(2012)}]{ISO-NE}%
  \BibitemOpen
  \href@noop {} {\enquote {\bibinfo {title} {{I}{S}{O} {New} {England}:
  Adressing gas dependence,
  \url{http://www.iso-ne.com/committees/comm_wkgrps/strategic_planning_discussion/materials/natural-gas-white-paper-draft-july-2012.pdf}},}\
  } (\bibinfo {year} {2012})\BibitemShut {NoStop}%
\bibitem [{\citenamefont {Kost}\ and\ \citenamefont
  {et~al}()}]{cost_renewables}%
  \BibitemOpen
  \bibfield  {author} {\bibinfo {author} {\bibfnamefont {C.}~\bibnamefont
  {Kost}}\ and\ \bibinfo {author} {\bibnamefont {et~al}},\ }\href@noop {}
  {\emph {\bibinfo {title} {Levelized cost of electricity renewable energy
  technologies}}},\ \bibinfo {type} {Tech. Rep.}\BibitemShut {Stop}%
\bibitem [{\citenamefont {Cory}\ and\ \citenamefont
  {Swezey}(2007)}]{portfolio_renewables}%
  \BibitemOpen
  \bibfield  {author} {\bibinfo {author} {\bibfnamefont {K.}~\bibnamefont
  {Cory}}\ and\ \bibinfo {author} {\bibfnamefont {B.}~\bibnamefont {Swezey}},\
  }\href@noop {} {\emph {\bibinfo {title} {Renewable Portfolio Standards in the
  States: Balancing Goals and Implementation Strategies}}},\ \bibinfo {type}
  {Tech. Rep.}\ (\bibinfo {year} {2007})\BibitemShut {NoStop}%
\bibitem [{201(2010)}]{2010MITEI}%
  \BibitemOpen
  \href@noop {} {\enquote {\bibinfo {title} {The future of natural gas:mit
  energy initiative,
  \url{http://mitei.mit.edu/system/files/NaturalGas_Report.pdf}},}\ } (\bibinfo
  {year} {2010})\BibitemShut {NoStop}%
\bibitem [{201()}]{2013MITEI}%
  \BibitemOpen
  \href@noop {} {\enquote {\bibinfo {title} {Growing concerns, possible
  solutions: The interdependency of natural gas and electricity systems},}\
  }\BibitemShut {NoStop}%
\bibitem [{\citenamefont {Osiadacz}()}]{osiadacz1987simulation}%
  \BibitemOpen
  \bibfield  {author} {\bibinfo {author} {\bibfnamefont {A.}~\bibnamefont
  {Osiadacz}},\ }\href@noop {} {\emph {\bibinfo {title} {Simulation and
  analysis of gas networks}}}\BibitemShut {NoStop}%
\bibitem [{\citenamefont {Thorley}\ and\ \citenamefont {Tiley}(1987)}]{87TT}%
  \BibitemOpen
  \bibfield  {author} {\bibinfo {author} {\bibfnamefont {A.}~\bibnamefont
  {Thorley}}\ and\ \bibinfo {author} {\bibfnamefont {C.}~\bibnamefont
  {Tiley}},\ }\href@noop {} {\bibfield  {journal} {\bibinfo  {journal}
  {International Journal of Heat and Fluid Flow}\ }\textbf {\bibinfo {volume}
  {8}},\ \bibinfo {pages} {3 } (\bibinfo {year} {1987})}\BibitemShut {NoStop}%
\bibitem [{\citenamefont {Sardanashvili}(2005)}]{05Sar}%
  \BibitemOpen
  \bibfield  {author} {\bibinfo {author} {\bibfnamefont {S.~A.}\ \bibnamefont
  {Sardanashvili}},\ }\href@noop {} {\emph {\bibinfo {title} {Computational
  Techniques and Algorithms (Pipeline Gas Transmission) [in Russian]}}}\
  (\bibinfo  {publisher} {FSUE "Oil and Gaz", I.M. Gubkin, Russian State
  University of Oil and Gas},\ \bibinfo {year} {2005})\BibitemShut {NoStop}%
\bibitem [{\citenamefont {Wong}\ and\ \citenamefont {Larson}(1968)}]{68WL}%
  \BibitemOpen
  \bibfield  {author} {\bibinfo {author} {\bibfnamefont {P.}~\bibnamefont
  {Wong}}\ and\ \bibinfo {author} {\bibfnamefont {R.}~\bibnamefont {Larson}},\
  }\href@noop {} {\bibfield  {journal} {\bibinfo  {journal} {Automatic Control,
  IEEE Transactions on}\ }\textbf {\bibinfo {volume} {13}},\ \bibinfo {pages}
  {475} (\bibinfo {year} {1968})}\BibitemShut {NoStop}%
\bibitem [{\citenamefont {Wu}\ \emph {et~al.}(2000)\citenamefont {Wu},
  \citenamefont {Ríos-Mercado}, \citenamefont {Boyd},\ and\ \citenamefont
  {Scott}}]{00WRBS}%
  \BibitemOpen
  \bibfield  {author} {\bibinfo {author} {\bibfnamefont {S.}~\bibnamefont
  {Wu}}, \bibinfo {author} {\bibfnamefont {R.}~\bibnamefont {Ríos-Mercado}},
  \bibinfo {author} {\bibfnamefont {E.}~\bibnamefont {Boyd}}, \ and\ \bibinfo
  {author} {\bibfnamefont {L.}~\bibnamefont {Scott}},\ }\href@noop {}
  {\bibfield  {journal} {\bibinfo  {journal} {Mathematical and Computer
  Modelling}\ }\textbf {\bibinfo {volume} {31}},\ \bibinfo {pages} {197 }
  (\bibinfo {year} {2000})}\BibitemShut {NoStop}%
\bibitem [{\citenamefont {Babonneau}\ \emph {et~al.}(2012)\citenamefont
  {Babonneau}, \citenamefont {Nesterov},\ and\ \citenamefont {Vial}}]{12BNV}%
  \BibitemOpen
  \bibfield  {author} {\bibinfo {author} {\bibfnamefont {F.}~\bibnamefont
  {Babonneau}}, \bibinfo {author} {\bibfnamefont {Y.}~\bibnamefont {Nesterov}},
  \ and\ \bibinfo {author} {\bibfnamefont {J.-P.}\ \bibnamefont {Vial}},\
  }\href@noop {} {\  (\bibinfo {year} {2012})}\BibitemShut {NoStop}%
\bibitem [{\citenamefont {Misra}\ \emph {et~al.}(2014)\citenamefont {Misra},
  \citenamefont {Fisher}, \citenamefont {Backhaus}, \citenamefont {Bent},
  \citenamefont {Chertkov},\ and\ \citenamefont {Pan}}]{13MFBBCP}%
  \BibitemOpen
  \bibfield  {author} {\bibinfo {author} {\bibfnamefont {S.}~\bibnamefont
  {Misra}}, \bibinfo {author} {\bibfnamefont {M.~W.}\ \bibnamefont {Fisher}},
  \bibinfo {author} {\bibfnamefont {S.}~\bibnamefont {Backhaus}}, \bibinfo
  {author} {\bibfnamefont {R.}~\bibnamefont {Bent}}, \bibinfo {author}
  {\bibfnamefont {M.}~\bibnamefont {Chertkov}}, \ and\ \bibinfo {author}
  {\bibfnamefont {F.}~\bibnamefont {Pan}},\ }\href@noop {} {\bibfield
  {journal} {\bibinfo  {journal} {IEEE Transactions on Control of Network
  Systems, to appear}\ } (\bibinfo {year} {2014})}\BibitemShut {NoStop}%
\bibitem [{Tra(2014)}]{Transco}%
  \BibitemOpen
  \href@noop {} {\enquote {\bibinfo {title} {The william transco pipe line,
  \url{http://www.1line.williams.com/Transco/index.html}},}\ } (\bibinfo {year}
  {2014})\BibitemShut {NoStop}%
\bibitem [{\citenamefont {Behbahani-Nejad}\ and\ \citenamefont
  {Shekari}(2010)}]{10BS}%
  \BibitemOpen
  \bibfield  {author} {\bibinfo {author} {\bibfnamefont {M.}~\bibnamefont
  {Behbahani-Nejad}}\ and\ \bibinfo {author} {\bibfnamefont {Y.}~\bibnamefont
  {Shekari}},\ }\href@noop {} {\bibfield  {journal} {\bibinfo  {journal} {J.
  Pet. Sci. Eng.}\ }\textbf {\bibinfo {volume} {73}},\ \bibinfo {pages} {13}
  (\bibinfo {year} {2010})}\BibitemShut {NoStop}%
\bibitem [{\citenamefont {Herty}\ \emph {et~al.}(2010)\citenamefont {Herty},
  \citenamefont {Mohring},\ and\ \citenamefont {Sachers}}]{11HMS}%
  \BibitemOpen
  \bibfield  {author} {\bibinfo {author} {\bibfnamefont {M.}~\bibnamefont
  {Herty}}, \bibinfo {author} {\bibfnamefont {J.}~\bibnamefont {Mohring}}, \
  and\ \bibinfo {author} {\bibfnamefont {V.}~\bibnamefont {Sachers}},\
  }\href@noop {} {\bibfield  {journal} {\bibinfo  {journal} {Mathematical
  Methods in the Applied Sciences}\ }\textbf {\bibinfo {volume} {33}},\
  \bibinfo {pages} {845} (\bibinfo {year} {2010})}\BibitemShut {NoStop}%
\bibitem [{\citenamefont {{CRANE}}(1982)}]{Ref_Crane1982}%
  \BibitemOpen
  \bibfield  {author} {\bibinfo {author} {\bibnamefont {{CRANE}}},\ }\href@noop
  {} {\emph {\bibinfo {title} {Flow of Fluids: Through Valves, Fittings and
  Pipe}}},\ \bibinfo {type} {Technical paper}\ \bibinfo {number} {410M}\
  (\bibinfo  {institution} {Crane Company},\ \bibinfo {address} {New York},\
  \bibinfo {year} {1982})\BibitemShut {NoStop}%
\bibitem [{\citenamefont {Mokhatab}\ \emph {et~al.}(2006)\citenamefont
  {Mokhatab}, \citenamefont {Poe},\ and\ \citenamefont
  {Speight}}]{Ref_Mokhatab2006}%
  \BibitemOpen
  \bibfield  {author} {\bibinfo {author} {\bibfnamefont {S.}~\bibnamefont
  {Mokhatab}}, \bibinfo {author} {\bibfnamefont {W.~A.}\ \bibnamefont {Poe}}, \
  and\ \bibinfo {author} {\bibfnamefont {J.~G.}\ \bibnamefont {Speight}},\
  }\href@noop {} {\emph {\bibinfo {title} {Handbook of Natural Gas Transmission
  and Processing}}}\ (\bibinfo  {publisher} {Gulf Professional Publishing},\
  \bibinfo {address} {Houston},\ \bibinfo {year} {2006})\BibitemShut {NoStop}%
\bibitem [{\citenamefont {Borraz-S\'{a}nchez}(2010)}]{10Bor}%
  \BibitemOpen
  \bibfield  {author} {\bibinfo {author} {\bibfnamefont {C.}~\bibnamefont
  {Borraz-S\'{a}nchez}},\ }\emph {\bibinfo {title} {Optimization Methods for
  Pipeline Transportation of Natural Gas}},\ \href@noop {} {Ph.D. thesis},\
  \bibinfo  {school} {Department of Informatics, University of Bergen, Norway}
  (\bibinfo {year} {2010})\BibitemShut {NoStop}%
\end{thebibliography}%

\end{document}